\documentstyle[12pt]{article}

\begin{document}

\title{Vacuum Expectation Values of the Quantum Fields}

\author{Yury M. Zinoviev\thanks{This work was supported in part by the Russian Foundation
 for Basic Research (Grant No. 07 - 01 - 00144) and Scientific Schools 672.2006.1}}

\date{}
\maketitle

\vskip 1cm

Steklov Mathematical Institute, Gubkin St. 8, 119991, Moscow, Russia,

 e - mail: zinoviev@mi.ras.ru

\vskip 1cm

\noindent {\bf Abstract.} The new axiomatic system for the quantum
field theory is proposed. The new axioms are the description of the
distributions. For the finite series these distributions satisfy the
linear Wightman axioms.

\vskip 2cm

\section{Introduction}
\setcounter{equation}{0}

G\aa rding -- Wightman axioms \cite{1} are the mathematical form of
the physical views of the quantum fields. Let us describe the finite
dimensional irreducible representations of the group $SL(2,{\bf
C})$, consisting of the complex $2\times 2$ - matrices
\begin{equation}
\label{1.1}
A =  \left( \begin{array}{cc}

A_{11} & A_{12} \\

A_{21} & A_{22}

\end{array} \right)
\end{equation}
with the determinant equal to $1$. The complex conjugate of the matrix (1.1) is
\begin{equation}
\label{1.2}
\bar{A} = \left( \begin{array}{cc}

\bar{A}_{11} & \bar{A}_{12} \\

\bar{A}_{21} & \bar{A}_{22}

\end{array} \right).
\end{equation}
Let us consider the values $\xi_{\alpha_{1} \cdots \alpha_{j} \dot{\beta}_{1} \cdots
\dot{\beta}_{k}}$ with the indices $\alpha$ and $\dot{\beta}$ equal to $1,2$. The
value $\xi$ is symmetric under the permutations of the indices $\alpha$ and the
permutations of the indices $\dot{\beta}$. For any matrix (\ref{1.1}) we define the
linear transformation
\begin{equation}
\label{1.3}
\sum_{(\rho )(\dot{\sigma} )} \left( \prod_{s = 1}^{j} A_{\alpha_{s}
\rho_{s}}\right) \left( \prod_{t = 1}^{k} \bar{A}_{\dot{\beta}_{t} \dot{\sigma}_{t}}
\right) \xi_{\rho_{1} \cdots \rho_{j} \dot{\sigma}_{1} \cdots \dot{\sigma}_{k}}.
\end{equation}
The dot over the index means simply that the value with this index
transforms with a matrix $\bar{A}$; the symbol $(\rho)$ means
$\rho_{1} \cdots \rho_{j}$; the symbol $(\dot{\sigma})$ means
$\dot{\sigma}_{1} \cdots \dot{\sigma}_{k}$. This representation of
the group $SL(2, {\bf C})$ is denoted by $D^{(j/2,k/2)}(A,\bar{A})$.
Any finite dimensional irreducible representation of the group
$SL(2, {\bf C})$ is equivalent to one of these representations.

Due to the book \cite{2} we formulate the properties of the vacuum expectation
values of the products of the quantum fields.

\noindent {\bf Spectrality}

\noindent {\it The vacuum expectation value of the product of} $n +
1, n = 1,2,...$, {\it quantum field is the distribution}
\begin{eqnarray}
\label{1.4} W_{(\alpha) (\dot{\beta}) ,...,(\gamma)
(\dot{\delta})}(x_{2} - x_{1},...,x_{n + 1} - x_{n}) = \nonumber \\
\int d^{4n}p \widetilde{W}_{(\alpha) (\dot{\beta}) ,...,(\gamma)
(\dot{\delta})} (p_{1},...,p_{n}) \exp \{ i\sum_{j = 1}^{n}
(p_{j},x_{j + 1} - x_{j})\},
\end{eqnarray}
\begin{equation}
\label{1.5}
(x,y) = x^{0}y^{0} - \sum_{k = 1}^{3} x^{k}y^{k}.
\end{equation}
{\it The distribution}
\begin{equation}
\label{1.6} \widetilde{W}_{(\alpha) (\dot{\beta}) ,...,(\gamma)
(\dot{\delta})} (p_{1},...,p_{n}) \in S^{\prime}({\bf R}^{4n})
\end{equation}
{\it has the support in the product of the closed upper light cones}
\begin{equation}
\label{1.7}
\overline{V}_{+} = \{ x \in {\bf R}^{4}: x^{0} \geq 0, (x,x) \geq 0 \}.
\end{equation}
The matrix
\begin{equation}
\label{1.8}
A^{T} = \left( \begin{array}{cc}

A_{11} & A_{21} \\

A_{12} & A_{22}

\end{array} \right)
\end{equation}
is called the transposed matrix. The matrix $A^{\ast} =
(\bar{A})^{T}$ is called Hermitian adjoint matrix. If $A^{\ast} =
A$, the matrix (\ref{1.1}) is Hermitian. Let us consider the basis
of the Hermitian $2\times 2$ - matrices
\begin{equation}
\label{1.9}
\sigma^{0} = \left( \begin{array}{cc}

1 & 0 \\

0 & 1

\end{array} \right),
\sigma^{1} = \left( \begin{array}{cc}

0 & 1 \\

1 & 0

\end{array} \right),
\sigma^{2} = \left( \begin{array}{cc}

0 & - i \\

i &   0

\end{array} \right),
\sigma^{3} = \left( \begin{array}{cc}

1 & 0 \\

0 & - 1

\end{array} \right).
\end{equation}
We relate with a vector $x \in {\bf R}^{4}$ an Hermitian matrix
\begin{equation}
\label{1.10}
\tilde{x} = \sum_{\mu = 0}^{3} x^{\mu} \sigma^{\mu}.
\end{equation}
We identify a vector $x \in {\bf R}^{4}$ and an Hermitian matrix (\ref{1.10}).

\noindent {\bf Lorentz covariance}

\noindent {\it For any matrix} (\ref{1.1})
\begin{eqnarray}
\label{1.11} \widetilde{W}_{(\alpha) (\dot{\beta}) ,...,(\gamma)
(\dot{\delta})} (A\tilde{p}_{1} A^{\ast},...,A\tilde{p}_{n}
A^{\ast}) = \nonumber \\ \sum_{(\alpha^{\prime})
(\dot{\beta^{\prime}}),..., (\gamma^{\prime})
(\dot{\delta^{\prime}})} D_{(\alpha) (\dot{\beta}),(\alpha^{\prime})
(\dot{\beta^{\prime}})}^{(j_{1}/2,k_{1}/2)} (A,\bar{A} ) \cdots
D_{(\gamma) (\dot{\delta}),(\gamma^{\prime})
(\dot{\delta^{\prime}})}^{(j_{n + 1}/2,k_{n + 1}/2)} (A,\bar{A} )
\widetilde{W}_{(\alpha^{\prime}) (\dot{\beta^{\prime}})
,...,(\gamma^{\prime}) (\dot{\delta^{\prime}})} (\tilde{p}_{1}
,...,\tilde{p}_{n}).
\end{eqnarray}
{\bf Locality}

\noindent {\it Let} $\pi$ {\it be a permutation of the natural
numbers} $1,...,n + 1$. {\it If} $(x_{i} - x_{j},x_{i} - x_{j}) < 0$
{\it for all numbers} $i,j$ {\it whose order is changed by a
permutation} $\pi$, {\it then}
\begin{eqnarray}
\label{1.12}
W_{\pi; (\alpha) (\dot{\beta}) ,...,(\gamma) (\dot{\delta})}(x_{\pi (2)} - x_{\pi
(1)},...,x_{\pi (n + 1)} - x_{\pi (n)}) = \nonumber \\ (- 1)^{M} W_{(\alpha)
(\dot{\beta}) ,...,(\gamma) (\dot{\delta})}(x_{2} - x_{1},...,x_{n + 1} - x_{n})
\end{eqnarray}
{\it where} $M$ {\it is the number of the transpositions of the anti - commuting
fields under the permutation} $\pi$. {\it In general case the distribution in the
left - hand side of the equality} (\ref{1.12}) {\it is another vacuum expectation
value of the product of the quantum fields. This vacuum expectation value depends on
the permutation} $\pi$.

Lorentz invariant distributions are studied in the papers \cite{3} -
\cite{5}. The description of Lorentz covariant tempered
distributions with supports in the product of the closed upper light
cones (\ref{1.7}) is obtained in the paper \cite{6}. The attempts to
construct a non - trivial example of the distributions satisfying
the properties of spectrality, Lorentz covariance and locality
failed.

The aim of this paper is the mathematical statement of the
properties of the vacuum expectation values of the quantum fields
products. The distributions (\ref{1.4}) for $n = 1$ are described in
the paper \cite{6}. We suppose that the distributions (\ref{1.4})
for $n > 1$ are similar to the distributions (\ref{1.4}) for $n =
1$. Our requirements are connected with three groups: the group of
the translations, Lorentz group and the group of the permutations of
the natural numbers $1,...,n + 1$. The vacuum expectation values of
the products of the quantum fields are the series of the
distributions. If these series are finite, the vacuum expectation
values of the products of the quantum fields satisfy the properties
of spectrality, Lorentz covariance and locality. We define the
convergence of the infinite series of the distributions in such a
manner that the vacuum expectation values of the products of the
quantum fields satisfy the properties of Lorentz covariance and
locality. We propose a new asymptotic condition for the vacuum
expectation values of the products of the quantum fields. We deal
with the asymptotic quantum fields in an experiment. The asymptotic
quantum fields may be considered free. The positivity
property~\cite{2} should be applied to the asymptotic vacuum
expectation values of the products of the quantum fields.

\section{Lorentz covariance and spectral condition}
\setcounter{equation}{0}

The group $SU(2)$ is the maximal compact subgroup of the group
$SL(2,{\bf C})$. It consists of the matrices from the group
$SL(2,{\bf C})$ satisfying the equation $AA^{\ast} = \sigma^{0}$.
Let us describe the irreducible representations of the group
$SU(2)$. We consider the half - integers $l \in 1/2{\bf Z}_{+}$,
i.e. $l = 0,1/2,1,3/2,...$. We define the representation of the
group $SU(2)$ on the space of the polynomials with degree less than
or equal to $2l$
\begin{equation}
\label{2.1} T_{l}(A)\phi (z) = (A_{12}z + A_{22})^{2l}\phi \left(
 \frac{A_{11}z + A_{21}}{A_{12}z + A_{22}}\right).
\end{equation}
We consider a half - integer $n = - l, - l + 1,...,l - 1,l$ and
choose the polynomial basis
\begin{equation}
\label{2.2} \psi_{n} (z) = ((l - n)!(l + n)!)^{- 1/2}z^{l - n}.
\end{equation}
The definitions (\ref{2.1}), (\ref{2.2}) imply
\begin{equation}
\label{2.3}
T_{l}(A)\psi_{n} (z) = \sum_{m = - l}^{l} \psi_{m} (z)t_{mn}^{l}(A)
\end{equation}
where the polynomial
\begin{eqnarray}
\label{2.4}
t_{mn}^{l}(A) = ((l - m)!(l + m)!(l - n)!(l + n)!)^{1/2} \times \nonumber \\ \sum_{j
= - \infty}^{\infty} \frac{A_{11}^{l - m - j}A_{12}^{j}A_{21}^{m - n + j}A_{22}^{l +
n - j}}{\Gamma (j + 1)\Gamma (l - m - j + 1)\Gamma (m - n + j + 1)\Gamma (l + n - j
+ 1)}
\end{eqnarray}
and $\Gamma (z)$ is the gamma - function. The function $(\Gamma
(z))^{- 1}$ equals zero for $z = 0 , - 1, - 2,...$. Therefore the
series (\ref{2.4}) is a polynomial.

The relation (\ref{2.1}) defines a representation of the group $SU(2)$. Thus the
polynomial (\ref{2.4}) defines a representation of the group $SU(2)$
\begin{equation}
\label{2.5}
t_{mn}^{l}(AB) = \sum_{k = - l}^{l} t_{mk}^{l}(A)t_{kn}^{l}(B).
\end{equation}
This $(2l + 1)$ - dimensional representation is irreducible (\cite{7}, Chapter III,
Section 2.3).

The relations (\ref{2.4}), (\ref{2.5}) have an analytic continuation to the matrices
from the group $SL(2,{\bf C})$. By making the change $j \rightarrow j + n - m$ of
the summation variable in the equality (\ref{2.4}) we have
\begin{equation}
\label{2.6}
t_{mn}^{l}(A) = t_{nm}^{l}(A^{T}).
\end{equation}
Due to (\cite{7}, Chapter III, Section 8.3) we have
\begin{eqnarray}
\label{2.7}
t_{m_{1}n_{1}}^{l_{1}}(A)t_{m_{2}n_{2}}^{l_{2}}(A) = \sum_{l_{3} \in 1/2{\bf Z}_{+}}
\sum_{m_{3},n_{3} = - l_{3}}^{l_{3}} \nonumber \\
C(l_{1},l_{2},l_{3};m_{1},m_{2},m_{3})C(l_{1},l_{2},l_{3};n_{1},n_{2},n_{3})t_{m_{3}n_{3}}^{l_{3}}(A)
\end{eqnarray}
for a matrix $A \in SU(2)$. The Clebsch - Gordan coefficient
$C(l_{1},l_{2},l_{3};m_{1},m_{2},m_{3})$ is not zero only if $m_{3}
= m_{1} + m_{2}$ and the half - integers $l_{1},l_{2},l_{3} \in
1/2{\bf Z}_{+}$ satisfy the triangle condition: it is possible to
construct a triangle with the sides of length $l_{1},l_{2},l_{3}$
and an integer perimeter $l_{1} + l_{2} + l_{3}$. It means that the
half - integer $l_{3}$ is one of the half - integers $|l_{1} -
l_{2}|,|l_{1} - l_{2}| + 1,...,l_{1} + l_{2} - 1,l_{1} + l_{2}$. Let
the half - integers $l_{1},l_{2},l_{3} \in 1/2{\bf Z}_{+}$ satisfy
the triangle condition. Let the half - integers $m_{i} = - l_{i}, -
l_{i} + 1,...,l_{i} - 1,l_{i}, i = 1,2,3$, $m_{3} = m_{1} + m_{2}$.
Then due to (\cite{7}, Chapter III, Section 8.3)
\begin{eqnarray}
\label{2.8}
C(l_{1},l_{2},l_{3};m_{1},m_{2},m_{3}) = (- 1)^{l_{1} - l_{3} + m_{2}} (2l_{3} +
1)^{1/2} \times \nonumber \\
 \left( \frac{(l_{1} + l_{2} - l_{3})!(l_{1} + l_{3} - l_{2})!(l_{2} + l_{3} -
l_{1})!(l_{3} - m_{3})!(l_{3} + m_{3})!}{(l_{1} + l_{2} + l_{3} + 1)!(l_{1} -
m_{1})!(l_{1} + m_{1})!(l_{2} - m_{2})!(l_{2} + m_{2})!}\right)^{1/2}
\times \nonumber \\
\sum_{j = 0}^{l_{2} + l_{3} - l_{1}} \frac{(- 1)^{j}(l_{1} + m_{1} + j)!)(l_{2} +
l_{3} - m_{1} - j)!}{j! \Gamma (l_{3} - m_{3} - j + 1) \Gamma (l_{1} - l_{2} + m_{3}
+ j + 1)(l_{2} + l_{3} - l_{1} - j)!}.
\end{eqnarray}
Let $dA$ be the normalized Haar measure on the group $SU(2)$. Due to (\cite{7},
Chapter III, Section 8.3)
\begin{eqnarray}
\label{2.9}
C(l_{1},l_{2},l_{3};m_{1},m_{2},m_{3})C(l_{1},l_{2},l_{3};n_{1},n_{2},n_{3}) =
\nonumber \\ (2l_{3} + 1)\int_{SU(2)} dA
t_{m_{1}n_{1}}^{l_{1}}(A)t_{m_{2}n_{2}}^{l_{2}}(A)\overline{t_{m_{3}n_{3}}^{l_{3}}(A)}.
\end{eqnarray}
The coefficients of the polynomial (\ref{2.4}) are real. By using the relations
(\ref{2.6}) and $A^{\ast} = A^{- 1}$ we can rewrite the equality (\ref{2.9}) as
\begin{eqnarray}
\label{2.10}
C(l_{1},l_{2},l_{3};m_{1},m_{2},m_{3})C(l_{1},l_{2},l_{3};n_{1},n_{2},n_{3}) =
\nonumber \\ (2l_{3} + 1)\int_{SU(2)} dA
t_{m_{1}n_{1}}^{l_{1}}(A)t_{m_{2}n_{2}}^{l_{2}}(A)t_{n_{3}m_{3}}^{l_{3}}(A^{- 1}).
\end{eqnarray}
If the half - integers $l_{1},l_{2},l_{3} \in 1/2{\bf Z}_{+}$ satisfy the triangle
condition, then due to (\cite{7}, Chapter III, Section 8.3) we have
\begin{equation}
\label{2.11} C(l_{1},l_{2},l_{3};l_{1},- l_{2},l_{1} - l_{2}) =
\left( \frac{(2l_{3} + 1)(2l_{1})!(2l_{2})!}{(l_{1} + l_{2} -
l_{3})!(l_{1} + l_{2} + l_{3} + 1)!}\right)^{1/2}.
\end{equation}
Let us choose the half - integers $n_{1} = l_{1}, n_{2} = - l_{2}$ in the equality
(\ref{2.10}). Then the relation (\ref{2.10}) and the invariance of the Haar measure
$dA$ imply
\begin{eqnarray}
\label{2.12}
\sum_{n_{1} = - l_{1}}^{l_{1}} \sum_{n_{2} = - l_{2}}^{l_{2}}
t_{m_{1}n_{1}}^{l_{1}}(A)t_{m_{2}n_{2}}^{l_{2}}(A)C(l_{1},l_{2},l_{3};n_{1},n_{2},m_{3})
= \nonumber \\ \sum_{n_{3} = - l_{3}}^{l_{3}}
C(l_{1},l_{2},l_{3};m_{1},m_{2},n_{3})t_{n_{3}m_{3}}^{l_{3}}(A).
\end{eqnarray}
The substitution of the matrix (\ref{1.8}) into the equality
(\ref{2.12}) and the equality (\ref{2.6}) yield
\begin{eqnarray}
\label{2.13}
\sum_{n_{1} = - l_{1}}^{l_{1}} \sum_{n_{2} = - l_{2}}^{l_{2}}
t_{n_{1}m_{1}}^{l_{1}}(A)t_{n_{2}m_{2}}^{l_{2}}(A)C(l_{1},l_{2},l_{3};n_{1},n_{2},m_{3})
= \nonumber \\ \sum_{n_{3} = - l_{3}}^{l_{3}}
C(l_{1},l_{2},l_{3};m_{1},m_{2},n_{3})t_{m_{3}n_{3}}^{l_{3}}(A).
\end{eqnarray}
The relations (\ref{2.12}) (\ref{2.13}) have an analytic continuation to any matrix
from the group $SL(2,{\bf C})$.

For any natural numbers $m,n$ and the half - integers
$l_{1},...,l_{n + 1},\dot{l}_{1},...,\dot{l}_{n + 1} \in 1/2{\bf
Z}_{+}; m_{i} = - l_{i}, - l_{i} + 1,...,l_{i} - 1,l_{i},
\dot{m}_{i} = - \dot{l}_{i}, - \dot{l}_{i} + 1,...,\dot{l}_{i} -
1,\dot{l}_{i}, i = 1,...,n + 1$, we consider the set of the tempered
distributions
$$
F_{m_{1},...,m_{n + 1};\dot{m}_{1},...,\dot{m}_{n + 1}}^{l_{1},...,l_{n +
1};\dot{l}_{1},...,\dot{l}_{n + 1}}(x_{1},...,x_{m}) \in S^{\prime}({\bf R}^{4m}).
$$
This set is called a Lorentz covariant distribution if for any matrix $A \in
SL(2,{\bf C})$
\begin{eqnarray}
\label{2.14}
F_{m_{1},...,m_{n + 1};\dot{m}_{1},...,\dot{m}_{n + 1}}^{l_{1},...,l_{n +
1};\dot{l}_{1},...,\dot{l}_{n + 1}}(A\tilde{x}_{1} A^{\ast},...,A\tilde{x}_{m}
A^{\ast}) = \sum_{k_{1} = - l_{1}}^{l_{1}} \cdots \sum_{k_{n + 1} = - l_{n +
1}}^{l_{n + 1}} \sum_{\dot{k}_{1} = - \dot{l}_{1}}^{\dot{l}_{1}} \cdots
\sum_{\dot{k}_{n + 1} = - \dot{l}_{n + 1}}^{\dot{l}_{n + 1}} \nonumber \\
\left( \prod_{i = 1}^{n + 1}
t_{m_{i}k_{i}}^{l_{i}}(A)t_{\dot{m}_{i}\dot{k}_{i}}^{\dot{l}_{i}}(\bar{A})
\right) F_{k_{1},...,k_{n + 1};\dot{k}_{1},...,\dot{k}_{n +
1}}^{l_{1},...,l_{n + 1};\dot{l}_{1},...,\dot{l}_{n +
1}}(\tilde{x}_{1},...,\tilde{x}_{m})
\end{eqnarray}
where $2\times 2$ - matrix $\tilde{x}$ is given by the relation (\ref{1.10}). The
half - integers $l_{1},...,l_{n + 1},\dot{l}_{1},...,\dot{l}_{n + 1}$ in the
relation (\ref{2.14}) are not arbitrary. Let us choose the matrix $A = - \sigma^{0}$
in the equality (\ref{2.14}). For this matrix
$$
A\tilde{x}_{j} A^{\ast} = \tilde{x}_{j}, j = 1,...,m.
$$
The definition (\ref{2.4}) implies
$$
t_{m_{j}k_{j}}^{l_{j}}(- \sigma^{0}) = (- 1)^{2l_{j}}\delta_{m_{j}k_{j}}.
$$
Hence the equality (\ref{2.14}) is valid for the matrix $A = - \sigma^{0}$ if
$$
(- 1)^{2(l_{1} + \cdots + l_{n + 1} + \dot{l}_{1} + \cdots + \dot{l}_{n + 1})} = 1
$$
i.e. the sum $l_{1} + \cdots + l_{n + 1} + \dot{l}_{1} + \cdots + \dot{l}_{n + 1}$
is an integer. This condition is supposed below.

Let a tempered distribution $F(x) \in S^{\prime }({\bf R}^{4})$ have a support in
the closed upper light cone (\ref{1.7}). Due to the paper \cite{8} there is a
natural number $q$ such that
\begin{equation}
\label{2.15}
F(x) = (\partial_{x}, \partial_{x})^{q} f(x),
\end{equation}
\begin{equation}
\label{2.16}
(\partial_{x}, \partial_{x}) = \left( \frac{\partial}{\partial x^{0}}\right)^{2} -
\sum_{k = 1}^{3} \left( \frac{\partial}{\partial x^{k}}\right)^{2}
\end{equation}
where a differentiable function $f(x)$ with a support in the closed upper light cone
(\ref{1.7}) is polynomial bounded.

Let us introduce the step function
\begin{equation}
\label{2.17} \theta (x) = \left\{ {1, \hskip 0,5cm x \geq 0,} \atop {0, \hskip 0,5cm x <
0.} \right.
\end{equation}
Due to (\cite{6}, Proposition) any tempered distribution
$$
F_{m_{1},m_{2};\dot{m}_{1}, \dot{m}_{2}}^{l_{1},l_{2};\dot{l}_{1}, \dot{l}_{2}}(x)
\in S^{\prime}({\bf R}^{4})
$$
with a support in the closed upper light cone (\ref{1.7}) satisfying the covariance
relation (\ref{2.14}) for $m = n = 1$ has the following form
\begin{eqnarray}
\label{2.18}
\int
d^{4}xF_{m_{1},m_{2};\dot{m}_{1},\dot{m}_{2}}^{l_{1},l_{2};\dot{l}_{1},\dot{l}_{2}}(x)\phi
(x) = \nonumber \\ \sum_{l_{3} \in 1/2{\bf Z}_{+}} \sum_{m_{3},\dot{m}_{3} = -
l_{3}}^{l_{3}}
C(l_{1},l_{2},l_{3};m_{1},m_{2},m_{3})C(\dot{l}_{1},\dot{l}_{2},l_{3};\dot{m}_{1},\dot{m}_{2},\dot{m}_{3})
\times \nonumber \\ \int d^{4}x\theta (x^{0})\theta
((x,x))t_{m_{3}\dot{m}_{3}}^{l_{3}}(\tilde{x})
f^{l_{1},l_{2},l_{3};\dot{l}_{1},\dot{l}_{2}}((x,x)^{1/2}) (\partial_{x},
\partial_{x})^{q}\phi (x)
\end{eqnarray}
where a test function $\phi (x) \in S({\bf R}^{4})$; $q$ is a natural number; the
Clebsch - Gordan coefficient $C(l_{1},l_{2},l_{3};m_{1},m_{2},m_{3})$ is given by
the relation (\ref{2.8}); $2\times 2$ - matrix $\tilde{x}$ is given by the relation
(\ref{1.10}); the polynomial $t_{mn}^{l}(A)$ is given by the relation (\ref{2.4});
the differentiable function $f^{l_{1},l_{2},l_{3};\dot{l}_{1},\dot{l}_{2}}(s)$ with
a support in the positive semi - axis is polynomial bounded.

Let us consider a tempered distribution with a support in the cone (\ref{1.7})
satisfying the covariance relation (\ref{2.14}) for $m = 1$, $n = 0$. This case
corresponds to the equality (\ref{2.18}) for $l_{2} = \dot{l}_{2} = 0$, $m_{2} =
\dot{m}_{2} = 0$. The relation (\ref{2.8}) implies
$$
C(l_{1},0,l_{3};m_{1},0,m_{3}) = \delta_{l_{1}l_{3}} \delta_{m_{1}m_{3}}.
$$

The vacuum expectation value of the product of two quantum field is
the Fourier transform of the distribution (\ref{2.18}). Let us
represent the distribution (\ref{2.18}) in the form suitable for
taking Fourier transform.

By making use of the matrices (\ref{1.9}) as the coefficients we define $2\times 2$
- matrix
\begin{equation}
\label{2.19}
 \tilde{\partial}_{x} = \sum_{\mu = 0}^{3} \eta^{\mu \mu} \sigma^{\mu} \frac{\partial}{\partial x^{\mu}}
\end{equation}
where $\eta^{00} = - \eta^{11} = - \eta^{22} = - \eta^{33} = 1$. We insert the
matrix (\ref{2.19}) into the polynomial (\ref{2.4}) and obtain the differential
operator $t_{mn}^{l}(\tilde{\partial}_{x} )$.

\noindent {\bf Lemma 2.1.} {\it Let} $f(s)$ {\it be} $2l$ {\it times differentiable
function. Then}
\begin{equation}
\label{2.20}
t_{mn}^{l}(\tilde{\partial}_{x} )f((x,x)) = t_{mn}^{l}(\tilde{x})
 \left( 2\frac{d}{ds} \right)^{2l} f(s)\bigr|_{s = (x,x)}
\end{equation}

\noindent {\it Proof.} The definitions (\ref{1.9}), (\ref{2.4}), (\ref{2.19}) imply
\begin{eqnarray}
\label{2.21}
t_{mn}^{l}(\tilde{\partial}_{x} )f((x,x)) = ((l - m)!(l + m)!(l - n)!(l + n)!)^{1/2}
\times \nonumber \\ \sum_{j = - \infty}^{\infty}
 \left( \Gamma (j + 1)\Gamma (l - m - j + 1)\Gamma (m - n + j + 1)\Gamma (l + n - j + 1)\right)^{- 1}
 \times \nonumber \\
\left( \frac{\partial}{\partial x^{0}} - \frac{\partial}{\partial x^{3}} \right)^{l
- m - j}
 \left( - \frac{\partial}{\partial x^{1}} + i\frac{\partial}{\partial x^{2}} \right)^{j}
 \times \nonumber \\ \left( (x^{1} + ix^{2})^{m - n + j}(x^{0} - x^{3})^{l + n - j}
 \left( 2\frac{d}{ds} \right)^{l + m} f(s)\bigr|_{s = (x,x)} \right).
\end{eqnarray}
By using Leibniz rule we have
\begin{eqnarray}
\label{2.22}
 \left( \frac{\partial}{\partial x^{0}} - \frac{\partial}{\partial x^{3}} \right)^{l - m - j}
 \left( - \frac{\partial}{\partial x^{1}} + i\frac{\partial}{\partial x^{2}} \right)^{j}
 \times \nonumber \\
 \left((x^{1} + ix^{2})^{m - n + j}(x^{0} - x^{3})^{l + n - j}
 \left( 2\frac{d}{ds} \right)^{l + m} f(s)\bigr|_{s =(x,x)} \right) = \nonumber \\
 \sum_{p = - \infty}^{\infty} \sum_{q = - \infty}^{\infty}
 \frac{\Gamma (l - m - j + 1)\Gamma (j + 1)}{\Gamma (p + 1)
 \Gamma (l - m - j - p + 1)\Gamma (q + 1)\Gamma (j - q + 1)} \times \nonumber \\
 \left( \left( \frac{\partial}{\partial x^{0}} -\frac{\partial}{\partial x^{3}} \right)^{p}
 (x^{0} - x^{3})^{l + n - j} \right)
 \left( \left( - \frac{\partial}{\partial x^{1}} + i\frac{\partial}{\partial x^{2}} \right)^{q}
 (x^{1} + ix^{2})^{m - n + j}\right) \times \nonumber \\
 \left( \frac{\partial}{\partial x^{0}} - \frac{\partial}{\partial x^{3}} \right)^{l - m - j - p}
 \left( - \frac{\partial}{\partial x^{1}} + i\frac{\partial}{\partial x^{2}} \right)^{j - q}
 \left( 2\frac{d}{ds} \right)^{l + m} f(s)\bigr|_{s = (x,x)}.
\end{eqnarray}
By calculating the derivatives in the right - hand side of the equality (\ref{2.22})
we can rewrite the equality (\ref{2.21}) in the form
\begin{eqnarray}
\label{2.23}
t_{mn}^{l}(\tilde{\partial}_{x} )f((x,x)) = ((l - m)!(l + m)!(l - n)!(l + n)!)^{1/2}
\times \nonumber \\ \sum_{j = - \infty}^{\infty} \sum_{p = - \infty}^{\infty}
\sum_{q = - \infty}^{\infty} (- 1)^{q}2^{p + q}
 \left( \Gamma (p + 1) \Gamma (q + 1) \right)^{- 1} \times \nonumber \\
 \left( \Gamma (j - q + 1) \Gamma (l - m - j - p + 1)
 \Gamma (m - n + j - q + 1) \Gamma (l + n - j  - p + 1)\right)^{- 1}
 \times \nonumber \\
 (x^{0} + x^{3})^{l - m - j - p}(x^{0} - x^{3})^{l + n - j - p}(x^{1} + ix^{2})^{m - n + j - q}
 (x^{1} - ix^{2})^{j - q} \times \nonumber \\
 \left( 2\frac{d}{ds} \right)^{2l - p - q} f(s)\bigr|_{s = (x,x)}.
\end{eqnarray}
By making the changes $j \rightarrow j + q$, $p \rightarrow p - q$ of the summation
variables we have
\begin{eqnarray}
\label{2.24}
t_{mn}^{l}(\tilde{\partial}_{x} )f((x,x)) = ((l - m)!(l + m)!(l - n)!(l + n)!)^{1/2}
\times \nonumber \\ \sum_{j = - \infty}^{\infty} \sum_{p = - \infty}^{\infty}
\sum_{q = - \infty}^{\infty} (- 1)^{q}2^{p}
 \left( \Gamma (p - q + 1) \Gamma (q + 1) \right)^{- 1} \times \nonumber \\
 \left( \Gamma (j + 1) \Gamma (l - m - j - p + 1)
 \Gamma (m - n + j + 1) \Gamma (l + n - j  - p + 1)\right)^{- 1}
 \times \nonumber \\
 (x^{0} + x^{3})^{l - m - j - p}(x^{0} - x^{3})^{l + n - j - p}(x^{1} + ix^{2})^{m - n + j}
 (x^{1} - ix^{2})^{j}
 \left( 2\frac{d}{ds} \right)^{2l - p} f(s)\bigr|_{s = (x,x)}.
\end{eqnarray}
The identity
\begin{equation}
\label{2.25}
\sum_{q = 0}^{p} (- 1)^{q} \frac{\Gamma (p + 1)}{ \Gamma (p - q + 1) \Gamma (q + 1)}
=
 (1 - 1)^{p}
\end{equation}
is valid. For $p \geq 1$ both sides of the equality (\ref{2.25}) are
equal to zero. For $p = 0$ the left - hand side of the equality
(\ref{2.25}) is equal to $1$. By making use of the equality
(\ref{2.25}) we can rewrite the equality (\ref{2.24}) in the form
(\ref{2.20}). The lemma is proved.

Let us define
\begin{equation}
\label{2.26} f^{( - 2l)}(s) = \theta (s) \int_{0}^{s} dt\frac{(s -
t)^{2l - 1}}{(2l - 1)!} f(t^{1/2}),\, \, 2l \geq 1,\, \,
 f^{(0)}(s) = \theta (s)f(s^{1/2}).
\end{equation}
By using the definitions (\ref{2.26}) we have
\begin{equation}
\label{2.27}
  \left( \frac{d}{ds}\right)^{2l} f^{( - 2l)}(s) = \theta (s)f(s^{1/2}).
\end{equation}
The equalities (\ref{2.20}), (\ref{2.27}) imply
\begin{equation}
\label{2.28}
  t_{mn}^{l}(\tilde{x}) \theta ((x,x))f((x,x)^{1/2}) =
  2^{- 2l}t_{mn}^{l}(\tilde{\partial_{x}}) f^{( - 2l)}((x,x)).
\end{equation}
If the function $f(s)$ is polynomial bounded, then the function (\ref{2.26}) is also
polynomial bounded. If the function $f(s)$ is continuous and has the support in the
positive semi - axis, then the function (\ref{2.26}) is also continuous and has the
support in the positive semi - axis. The function (\ref{2.26}) is $2l$ times
differentiable and its first $2l$ derivatives are equal to zero at the point $s =
0$. Therefore the equality (\ref{2.28}) implies
\begin{eqnarray}
\label{2.29}
  \theta (x^{0})\theta ((x,x))t_{m_{3}\dot{m}_{3}}^{l_{3}}(\tilde{x})
  f^{l_{1},l_{2},l_{3};\dot{l}_{1},\dot{l}_{2}}((x,x)^{1/2}) = \nonumber \\
  t_{m_{3}\dot{m}_{3}}^{l_{3}}(\tilde{\partial_{x}})
  \left( \theta (x^{0})2^{- 2l_{3}}\left( f^{l_{1},l_{2},l_{3};\dot{l}_{1}, \dot{l}_{2}}\right)^{( - 2l_{3})}
  ((x,x))\right).
\end{eqnarray}
The substitution of the equality (\ref{2.29}) into the relation
(\ref{2.18}) gives the following form of any tempered distribution
$$
F_{m_{1},m_{2};\dot{m}_{1}, \dot{m}_{2}}^{l_{1},l_{2};\dot{l}_{1}, \dot{l}_{2}}(x)
\in S^{\prime}({\bf R}^{4})
$$
with a support in the closed upper light cone (\ref{1.7}) satisfying
the covariance relation (\ref{2.14}) for $m = n = 1$
\begin{eqnarray}
\label{2.30}
\int
d^{4}xF_{m_{1},m_{2};\dot{m}_{1},\dot{m}_{2}}^{l_{1},l_{2};\dot{l}_{1},\dot{l}_{2}}(x)\phi
(x) = \nonumber \\ \sum_{l_{3} \in 1/2{\bf Z}_{+}} \sum_{m_{3},\dot{m}_{3} = -
l_{3}}^{l_{3}}
C(l_{1},l_{2},l_{3};m_{1},m_{2},m_{3})C(\dot{l}_{1},\dot{l}_{2},l_{3};\dot{m}_{1},\dot{m}_{2},\dot{m}_{3})
\times \nonumber \\ \int d^{4}x\theta (x^{0})2^{- 2l_{3}} \left(
f^{l_{1},l_{2},l_{3};\dot{l}_{1},\dot{l}_{2}}\right)^{( - 2l_{3})}((x,x))
(\partial_{x},
\partial_{x})^{q}t_{m_{3}\dot{m}_{3}}^{l_{3}}( - \tilde{\partial_{x}})\phi (x)
\end{eqnarray}
where a test function $\phi (x) \in S({\bf R}^{4})$; $q$ is a natural number; the
Clebsch - Gordan coefficient $C(l_{1},l_{2},l_{3};m_{1},m_{2},m_{3})$ is given by
the relation (\ref{2.8}); $2\times 2$ - matrix $\tilde{\partial_{x}}$ is given by
the relation (\ref{2.19}); the polynomial $t_{mn}^{l}(A)$ is given by the relation
(\ref{2.4}); the continuous function $\left(
f^{l_{1},l_{2},l_{3};\dot{l}_{1},\dot{l}_{2}}\right)^{( - 2l_{3})}(s)$ with a
support in the positive semi - axis is polynomial bounded.

Let us consider the tempered distributions of the form
\begin{eqnarray}
\label{2.31}
\int d^{4m}x
   F_{m_{1},...,m_{n + 1};\dot{m}_{1},..., \dot{m}_{n + 1}}^{l_{1},...,
   l_{n + 1};\dot{l}_{1},..., \dot{l}_{n + 1}}(x_{1},...,x_{m})
  \phi (x_{1},...,x_{m}) = \nonumber \\
  \sum_{l_{n + 2},...,l_{n + m + 1} \in 1/2{\bf Z}_{+}}
  \sum_{m_{s},\dot{m}_{s} \, \, =\, \, - l_{s}, - l_{s} + 1,...,l_{s} - 1,l_{s},\, \, s\, \, =\, \, n + 2,...,n + m + 1}
  \nonumber \\
  \int d^{4m}xf_{m_{1},...,m_{n + m + 1};\dot{m}_{1},...,
  \dot{m}_{n + m + 1}}^{l_{1},..., l_{n + m + 1};\dot{l}_{1},..., \dot{l}_{n + 1}}
  ((x_{1},x_{1}),...,(x_{m},x_{m})) \times \nonumber \\ \left( \prod_{j = 1}^{m} (\theta (x_{j}^{0})
  (\partial_{x_{j}}, \partial_{x_{j}})^{q}
  t_{m_{n + 1 + j}\dot{m}_{n + 1 + j}}^{l_{n + 1 + j}}( - \tilde{\partial}_{x_{j}})) \right)
  \phi (x_{1},...,x_{m})
\end{eqnarray}
where a test function $\phi (x_{1},...,x_{m}) \in S({\bf R}^{4m})$; $q$ is a natural
number; $2\times 2$ - matrix $\tilde{\partial}_{x}$ is given by the relation
(\ref{2.19}); the polynomial $t_{mn}^{l}(A)$ is given by the relation (\ref{2.4});
the continuous function
$$
f_{m_{1},...,m_{n + m + 1};\dot{m}_{1},...,
  \dot{m}_{n + m + 1}}^{l_{1},..., l_{n + m + 1};\dot{l}_{1},..., \dot{l}_{n + 1}} (s_{1},...,s_{m})
$$
with a support in the product of the positive semi - axes is polynomial bounded; the
sums over the summation variables $l_{n + 2},...,l_{n + m + 1}$ are finite. The
distribution (\ref{2.31}) belongs to the space $S^{\prime}({\bf R}^{4m})$. The sum
over the summation variable $l_{3}$ in the relation (\ref{2.30}) is finite due to
the triangle condition for the Clebsch - Gordan coefficients.

For any numbers $m,n \in {\bf Z}_{+}$ we define the generalized Clebsch - Gordan
coefficient
\begin{eqnarray}
\label{2.32}
C(l_{1},...,l_{m + 2};l_{m + 3},...,l_{m + n + 4};j_{1},...,j_{m + n + 1};
  m_{1},...,m_{m + 2};m_{m + 3},...,m_{m + n + 4}) = \nonumber \\
  \sum_{k_{s}\, \, =\, \, - j_{s}, - j_{s} + 1,...,j_{s} - 1,j_{s},\, \, s\, \, =\, \, 1,...,m + n + 1}
  C(l_{1},l_{2},j_{1};m_{1},m_{2},k_{1}) \times \nonumber \\
  \left( \prod_{s\, =\, 1}^{m} C(j_{s},l_{s + 2},j_{s + 1};k_{s},m_{s + 2},k_{s + 1}) \right)
  \left( \prod_{s\, =\, m + 1}^{m + n} C(j_{s + 1},l_{s + 2},j_{s};k_{s + 1},m_{s + 2},k_{s}) \right)
  \times \nonumber \\
  C(l_{m + n + 4},l_{m + n + 3},j_{m + n + 1};m_{m + n + 4},m_{m + n + 3},k_{m + n + 1})
\end{eqnarray}
where the half - integers $l_{1},...,l_{m + n + 4},j_{1},...,j_{m + n + 1} \in
1/2{\bf Z}_{+}$ and $m_{i} = - l_{i}, - l_{i} + 1,...,l_{i} - 1,l_{i}, i = 1,...,m +
n + 4$. The definition (\ref{2.32}) and the relations (\ref{2.12}), (\ref{2.13})
imply for any matrix $A \in SL(2,{\bf C})$
\begin{eqnarray}
\label{2.33}
\sum_{n_{s}\, \, =\, \, - l_{s}, - l_{s} + 1,...,l_{s} - 1,l_{s},\, \, s\, \, =\, \,
1,...,m + 2}
  \left( \prod_{i\, =\, 1}^{m + 2} t_{m_{i},n_{i}}^{l_{i}}(A)\right) \times
  \nonumber \\
C(l_{1},...,l_{m + 2};l_{m + 3},...,l_{m + n + 4};j_{1},...,j_{m + n + 1};
  n_{1},...,n_{m + 2};m_{m + 3},...,m_{m + n + 4}) = \nonumber \\
\sum_{n_{s}\, \, =\, \, - l_{s}, - l_{s} + 1,...,l_{s} - 1,l_{s},\, \, s\, \, =\, \,
m + 3,...,m + n + 4}
  \left( \prod_{i\, =\, m + 3}^{m + n + 4} t_{n_{i},m_{i}}^{l_{i}}(A)\right) \times
  \nonumber \\
C(l_{1},...,l_{m + 2};l_{m + 3},...,l_{m + n + 4};j_{1},...,j_{m + n + 1};
  m_{1},...,m_{m + 2};n_{m + 3},...,n_{m + n + 4}),
\end{eqnarray}
\begin{eqnarray}
\label{2.34}
\sum_{n_{s}\, \, =\, \, - l_{s}, - l_{s} + 1,...,l_{s} - 1,l_{s},\, \, s\, \, =\, \,
1,...,m + 2}
  \left( \prod_{i\, =\, 1}^{m + 2} t_{n_{i},m_{i}}^{l_{i}}(A)\right) \times
  \nonumber \\
C(l_{1},...,l_{m + 2};l_{m + 3},...,l_{m + n + 4};j_{1},...,j_{m + n + 1};
  n_{1},...,n_{m + 2};m_{m + 3},...,m_{m + n + 4}) = \nonumber \\
\sum_{n_{s}\, \, =\, \, - l_{s}, - l_{s} + 1,...,l_{s} - 1,l_{s},\, \, s\, \, =\, \,
m + 3,...,m + n + 4}
  \left( \prod_{i\, =\, m + 3}^{m + n + 4} t_{m_{i},n_{i}}^{l_{i}}(A)\right) \times
  \nonumber \\
C(l_{1},...,l_{m + 2};l_{m + 3},...,l_{m + n + 4};j_{1},...,j_{m + n + 1};
  m_{1},...,m_{m + 2};n_{m + 3},...,n_{m + n + 4}).
\end{eqnarray}
We will consider in the following sections the distributions
(\ref{2.31}) for the case $m = \frac{1}{2} n(n + 1)$ only.

\noindent {\bf Theorem 2.2.} {\it Let the distributions}
(\ref{2.31}) {\it for} $m = n = 1$ {\it satisfy the covariance
relation} (\ref{2.14}), {\it then}
\begin{equation}
\label{2.35}
f_{m_{1},m_{2},m_{3};\dot{m}_{1}, \dot{m}_{2},
\dot{m}_{3}}^{l_{1},l_{2},l_{3};\dot{l}_{1}, \dot{l}_{2}} (s_{1}) =
f^{l_{1},l_{2},l_{3};\dot{l}_{1}, \dot{l}_{2}} (s_{1})
  C(l_{1},l_{2},l_{3};m_{1},m_{2},m_{3})C(\dot{l}_{1}, \dot{l}_{2}, l_{3};\dot{m}_{1}, \dot{m}_{2}, \dot{m}_{3})
\end{equation}
{\it where a continuous function}
  $f^{l_{1}, l_{2}, l_{3}; \dot{l}_{1} ,\dot{l}_{2}} (s_{1})$ {\it with a support
  in the positive semi - axis is polynomial bounded; the
Clebsch - Gordan coefficient} $C(l_{1}, l_{2}, l_{3}; m_{1}, m_{2}, m_{3})$ {\it is
given by the relation} (\ref{2.8}).

{\it Let the distributions} (\ref{2.31}) {\it for} $m > 2, n > 1$ {\it satisfy the
covariance relation} (\ref{2.14}), {\it then}
\begin{eqnarray}
\label{2.36}
f_{m_{1},...,m_{n + m + 1};\dot{m}_{1},...,
  \dot{m}_{n + m + 1}}^{l_{1},..., l_{n + m + 1};\dot{l}_{1},..., \dot{l}_{n + 1}}
  (s_{1},...,s_{m}) = \nonumber \\
  \sum_{j_{1},...,j_{n + m - 2};j_{1}^{\prime},...,j_{n + m - 2}^{\prime} \in 1/2{\bf Z}_{+}}
  f_{j_{1},...,j_{n + m - 2};j_{1}^{\prime},...,
  j_{n + m - 2}^{\prime}}^{l_{1},..., l_{n + m + 1};\dot{l}_{1},..., \dot{l}_{n + 1}} (s_{1},...,s_{m})
  \times \nonumber \\
  C(l_{1},...,l_{n + 1};l_{n + 2},...,l_{m + n + 1};j_{1},...,j_{m + n - 2};
  m_{1},...,m_{n + 1};m_{n + 2},...,m_{m + n + 1}) \times \nonumber \\
  C(\dot{l}_{1},..., \dot{l}_{n + 1};l_{n + 2},...,l_{m + n + 1};j_{1}^{\prime},...,j_{m + n - 2}^{\prime};
  \dot{m}_{1},..., \dot{m}_{n + 1}; \dot{m}_{n + 2},..., \dot{m}_{m + n + 1})
\end{eqnarray}
{\it where a continuous function}
$$
f_{j_{1},...,j_{n + m - 2};j_{1}^{\prime},...,
  j_{n + m - 2}^{\prime}}^{l_{1},..., l_{n + m + 1};\dot{l}_{1},..., \dot{l}_{n + 1}} (s_{1},...,s_{m})
$$
{\it with a support in the product of the positive semi - axes is polynomial
bounded; the generalized Clebsch - Gordan coefficient}
$$
C(l_{1},...,l_{m + 2};l_{m + 3},...,l_{m + n + 4};j_{1},...,j_{m + n + 1};
  m_{1},...,m_{m + 2};m_{m + 3},...,m_{m + n + 4})
$$
{\it is given by the relation} (\ref{2.32}).

\noindent {\it Proof.} The covariance relation (\ref{2.14}) may be rewritten as
\begin{eqnarray}
\label{2.37}
F_{m_{1},...,m_{n + 1};\dot{m}_{1},..., \dot{m}_{n + 1}}^{l_{1},...,
   l_{n + 1};\dot{l}_{1},..., \dot{l}_{n + 1}}(\tilde{x}_{1},..., \tilde{x}_{m}) =
   \nonumber \\
   \sum_{k_{s}\,  = \, - l_{s}, - l_{s} + 1,...,l_{s} - 1,l_{s}, \, \, s\, =\, 1,...,n + 1 }
   \sum_{\dot{k}_{s}\,  = \, - \dot{l}_{s}, - \dot{l}_{s} + 1,...,\dot{l}_{s} - 1,\dot{l}_{s}, \, \, s\, =\, 1,...,n + 1 }
   \nonumber \\
\left( \prod_{i\, =\, 1}^{n + 1} t_{m_{i}k_{i}}^{l_{i}} (A^{- 1})t_{\dot{m}_{i}
\dot{k}_{i}}^{\dot{l}_{i}} (\bar{A}^{- 1})\right)
   F_{k_{1},...,k_{n + 1};\dot{k}_{1},..., \dot{k}_{n + 1}}^{l_{1},...,
   l_{n + 1};\dot{l}_{1},..., \dot{l}_{n + 1}}(A\tilde{x}_{1}A^{\ast},...,
   A\tilde{x}_{m}A^{\ast}).
\end{eqnarray}
Now the equality (\ref{2.31}) and the relations (\ref{2.5}), (\ref{2.6}) imply
\begin{eqnarray}
\label{2.38}
\int d^{4m}xF_{m_{1},...,m_{n + 1};\dot{m}_{1},..., \dot{m}_{n + 1}}^{l_{1},...,
   l_{n + 1};\dot{l}_{1},..., \dot{l}_{n + 1}}(\tilde{x}_{1},..., \tilde{x}_{m}) \phi (x_{1},...,x_{m}) =
   \nonumber \\
   \sum_{l_{n + 2},...,l_{n + m + 1} \in 1/2{\bf Z}_{+}}
   \sum_{k_{s}\,  = \, - l_{s}, - l_{s} + 1,...,l_{s} - 1,l_{s}, \, \, s\, =\, 1,...,n + 1 }
   \sum_{\dot{k}_{s}\,  = \, - \dot{l}_{s}, - \dot{l}_{s} + 1,...,\dot{l}_{s} - 1,\dot{l}_{s}, \, \, s\, =\, 1,...,n + 1 }
   \nonumber \\
   \sum_{m_{s}\dot{m}_{s}, k_{s},\dot{k}_{s} \,  = \, - l_{s}, - l_{s} + 1,...,l_{s} - 1,l_{s}, \, \,
   s\, =\, n + 2,...,n + m + 1 } \nonumber \\
   \left( \prod_{i\, =\, 1}^{n + 1} t_{m_{i}k_{i}}^{l_{i}} (A^{- 1})t_{\dot{m}_{i}
\dot{k}_{i}}^{\dot{l}_{i}} (\bar{A}^{- 1})\right)
   \left( \prod_{j\, =\, n + 2}^{n + m + 1} t_{m_{j}k_{j}}^{l_{j}} (A)t_{\dot{m}_{j}
\dot{k}_{j}}^{\dot{l}_{j}} (\bar{A})\right) \times \nonumber \\
   \int d^{4m}x f_{k_{1},...,k_{n + m + 1};\dot{k}_{1},..., \dot{k}_{n + m + 1}}^{l_{1},...,
   l_{n + m + 1};\dot{l}_{1},..., \dot{l}_{n + 1}}((x_{1},x_{1}),...,
   (x_{m},x_{m})) \times \nonumber \\
   \left( \prod_{j\, =\, 1}^{m} \theta (x_{j}^{0}) (\partial_{x_{j}}, \partial_{x_{j}})^{q}
   t_{k_{n + 1 + j},\dot{k}_{n + 1 + j}}^{l_{n + 1 + j}} (- \tilde{\partial}_{x_{j}})\right)
   \phi (x_{1},...,x_{m})
\end{eqnarray}
for any matrix $A \in SL(2,{\bf C})$.

The sums (\ref{2.38}) over the half - integers $l_{n + 2},...,l_{n +
m + 1}$ are finite. Hence the right - hand side of the equality
(\ref{2.38}) is a polynomial of the matrix elements of the matrices
$A,\bar{A}$. Let us choose the matrix from the group $SL(2, {\bf
C})$
\begin{eqnarray}
\label{2.39} A = \exp \{ \frac{1}{4} (\phi_{1}^{\prime} + i\phi_{1})
\sigma^{3} \} \exp \{ \frac{1}{4} (\theta_{1}^{\prime} +
i\theta_{1}) \sigma^{1} \} \exp \{ \frac{1}{4} (\psi_{1}^{\prime} +
i\psi_{1}) \sigma^{3} \} \times \nonumber \\ \exp \{ \frac{1}{4}
(\phi_{2}^{\prime} - i\phi_{2}) \sigma^{3} \} \exp \{ \frac{1}{4}
(\theta_{2}^{\prime} - i\theta_{2}) \sigma^{1} \} \exp \{
\frac{1}{4} (\psi_{2}^{\prime} - i\psi_{2}) \sigma^{3} \}.
\end{eqnarray}
Due to the definition (\ref{1.9}) the matrix elements of the
matrices $\sigma^{1}, \sigma^{3}$ are real. Therefore
\begin{eqnarray}
\label{2.40} \bar{A} = \exp \{ \frac{1}{4} (\phi_{1}^{\prime} -
i\phi_{1}) \sigma^{3} \} \exp \{ \frac{1}{4} (\theta_{1}^{\prime} -
i\theta_{1}) \sigma^{1} \} \exp \{ \frac{1}{4} (\psi_{1}^{\prime} -
i\psi_{1}) \sigma^{3} \} \times \nonumber \\ \exp \{ \frac{1}{4}
(\phi_{2}^{\prime} + i\phi_{2}) \sigma^{3} \} \exp \{ \frac{1}{4}
(\theta_{2}^{\prime} + i\theta_{2}) \sigma^{1} \} \exp \{
\frac{1}{4} (\psi_{2}^{\prime} + i\psi_{2}) \sigma^{3} \}.
\end{eqnarray}
The matrices (\ref{2.39}), (\ref{2.40}) have the analytic
continuation to the parameters $\phi_{k}^{\prime} = i\phi_{k},
\theta_{k}^{\prime} = \theta_{k} , \psi_{k}^{\prime} = i\psi_{k}, k
= 1,2$:
\begin{eqnarray}
\label{2.41} A = u(\phi_{1}, \theta_{1}, \psi_{1}), \bar{A} =
u(\phi_{2}, \theta_{2}, \psi_{2}), \nonumber \\
    u(\phi, \theta, \psi) = \exp \{ i\frac{\phi}{2} \sigma^{3} \} \exp \{ i\frac{\theta}{2} \sigma^{1} \} \exp \{ i\frac{\psi}{2} \sigma^{3} \}.
\end{eqnarray}
Due to the  (\cite{7}, Chapter III, Section 6.1) the normalized Haar
measure on the group $SU(2)$
\begin{equation}
\label{2.42} du(\phi, \theta, \psi) = (16\pi^{2})^{- 1} \sin \theta
d\theta d\phi d\psi.
\end{equation}
We insert the matrices (\ref{2.41}) into the relation (\ref{2.38})
and integrate the obtained relation over both groups $SU(2)$ with
the measures (\ref{2.42}).
\begin{eqnarray}
\label{2.43} \int d^{4m}xF_{m_{1},...,m_{n + 1};\dot{m}_{1},...,
\dot{m}_{n + 1}}^{l_{1},...,
   l_{n + 1};\dot{l}_{1},..., \dot{l}_{n + 1}}(\tilde{x}_{1},..., \tilde{x}_{m}) \phi (x_{1},...,x_{m}) =
   \nonumber \\
   \sum_{l_{n + 2},...,l_{n + m + 1} \in 1/2{\bf Z}_{+}}
   \sum_{k_{s}\,  = \, - l_{s}, - l_{s} + 1,...,l_{s} - 1,l_{s}, \, \, s\, =\, 1,...,n + 1 }
   \sum_{\dot{k}_{s}\,  = \, - \dot{l}_{s}, - \dot{l}_{s} + 1,...,\dot{l}_{s} - 1,\dot{l}_{s}, \, \, s\, =\, 1,...,n + 1 }
   \nonumber \\
   \sum_{m_{s}\dot{m}_{s}, k_{s},\dot{k}_{s} \,  = \, - l_{s}, - l_{s} + 1,...,l_{s} - 1,l_{s}, \, \,
   s\, =\, n + 2,...,n + m + 1 } \int_{SU(2)} du_{1} \int_{SU(2)} du_{2} \nonumber \\
   \left( \prod_{i\, =\, 1}^{n + 1} t_{m_{i}k_{i}}^{l_{i}} (u_{1}^{- 1})t_{\dot{m}_{i}
\dot{k}_{i}}^{\dot{l}_{i}} (u_{2}^{- 1})\right)
   \left( \prod_{j\, =\, n + 2}^{n + m + 1} t_{m_{j}k_{j}}^{l_{j}} (u_{1})t_{\dot{m}_{j}
\dot{k}_{j}}^{\dot{l}_{j}} (u_{2})\right) \times \nonumber \\
   \int d^{4m}x f_{k_{1},...,k_{n + m + 1};\dot{k}_{1},..., \dot{k}_{n + m + 1}}^{l_{1},...,
   l_{n + m + 1};\dot{l}_{1},..., \dot{l}_{n + 1}}((x_{1},x_{1}),...,
   (x_{m},x_{m})) \times \nonumber \\
   \left( \prod_{j\, =\, 1}^{m} \theta (x_{j}^{0}) (\partial_{x_{j}}, \partial_{x_{j}})^{q}
   t_{k_{n + 1 + j},\dot{k}_{n + 1 + j}}^{l_{n + 1 + j}} (- \tilde{\partial}_{x_{j}})\right)
   \phi (x_{1},...,x_{m})
\end{eqnarray}

Let us consider the case $m = n = 1$. In view of the definition of
the group $SU(2)$ we have $u^{- 1} = u^{\ast}$. Hence the relations
(\ref{2.6}), (\ref{2.9}), (\ref{2.43}) imply the equality
(\ref{2.35}). Due to the relations (\ref{2.12}), (\ref{2.13}) any
distribution (\ref{2.31}) with the functions (\ref{2.35}) satisfies
the covariance relation (\ref{2.14}).

Let us consider the case $m > 2$, $n > 1$. By making use of the
relations (\ref{2.6}), (\ref{2.7}), (\ref{2.9}) and the equality
(\ref{2.43}) we have the equality (\ref{2.36}). Due to the relations
(\ref{2.33}), (\ref{2.34}) any distribution (\ref{2.31}) with the
functions (\ref{2.36}) satisfies the covariance relation
(\ref{2.14}). The theorem is proved.

If we insert the functions (\ref{2.35}) into the series (\ref{2.31})
for $m = n = 1$, then this series will be finite due to the triangle
condition for the Clebsch - Gordan coefficients. If we insert  the
functions (\ref{2.36}) into the series (\ref{2.31}) for $m > 2$, $n
> 1$, then this series may be infinite. It is necessary to choose
the functions $\phi (x_{1},...,x_{m})$ for which the series is
convergent. If $\phi (x_{1},...,x_{m}) \in D({\bf R}^{4m})$, then
the infinitely differentiable function $\phi (x_{1},...,x_{m})$ has
a compact support. For these functions we can formulate the the
properties of spectrality and Lorentz covariance. We can not
formulate the property of locality for these functions. If the
Fourier transform
\begin{equation}
\label{2.44} \tilde{\phi} (p_{1},...,p_{m}) = \int d^{4m}x\exp \{
i\sum_{j\, =\, 1}^{m} (p_{j},x_{j}) \}
  \phi (x_{1},...,x_{m})
\end{equation}
belongs to the space $D({\bf R}^{4m})$, then the function
(\ref{2.44}) is infinitely differentiable and has a compact support.
For these functions we can formulate the properties of locality and
Lorentz covariance. We can not formulate the property of spectrality
for these functions.

Due to (\cite{9}, Section 30)
$$
(8\pi)^{- 1} (\partial_{x}, \partial_{x} )^{2} \left( \theta
(x^{0})\theta ((x,x))\right) = \delta (x),
$$
$$
(\partial_{x}, \partial_{x} )\left( \theta (x^{0})\theta ((x,x))
(x,x)^{n}\right) = 4n(n + 1)\theta (x^{0})\theta ((x,x))(x,x)^{n -
1}, n = 1,2,... .
$$
Let us consider the distribution (\ref{2.31}) for the case
\begin{eqnarray}
\label{2.45} \left( \prod_{j\, =\, 1}^{m} (\partial_{x_{j}},
\partial_{x_{j}})^{q}\right) \left( \left( \prod_{j\, =\, 1}^{m} \theta
(x^{0})\right) f_{m_{1},...,m_{n + m + 1};\dot{m}_{1},....,
\dot{m}_{n + m + 1}}^{l_{1},...,l_{n + m + 1};\dot{l}_{1},...,
\dot{l}_{n + m + 1}} ((x_{1},x_{1}),...,(x_{m},x_{m})) \right) =
\nonumber
\\ a_{m_{1},...,m_{n + m + 1};\dot{m}_{1},...., \dot{m}_{n + m +
1}}^{l_{1},...,l_{n + m + 1};\dot{l}_{1},..., \dot{l}_{n + m + 1}}
\prod_{j\, =\, 1}^{m} \delta (x_{j}).
\end{eqnarray}
The substitution of the equalities (\ref{2.45}) into the series
(\ref{2.31}) yields
\begin{equation}
\label{2.46} \sum_{k_{j}^{\mu} \, \in \, {\bf Z}_{+},\, \, j\, =\,
1,...,m,\, \, \mu \, =\, 0,...,3} a(k_{j}^{\mu}) \left( \prod_{j\,
=\, 1}^{m} \prod_{\mu \, =\, 0}^{3} \left( \frac{\partial}{\partial
x_{j}^{\mu}} \right)^{k_{j}^{\mu}} \right) \phi
(x_{1},...,x_{m})\bigr|_{x_{j}\, =\, 0}.
\end{equation}
{\bf Lemma 2.3.} {\it If the series} (\ref{2.46}) {\it is convergent
for any function} $\phi (x_{1},...,x_{m}) \in D({\bf R}^{4m})$, {\it
then the number of the coefficients } $a(k_{j}^{\mu})$ {\it which
are not equal to zero is finite.}

\noindent {\it Proof.} Let us prove that for any coefficients
$b(k_{j}^{\mu})$ there is the function $\phi (x_{1},...,x_{m}) \in
D({\bf R}^{4m})$ such that
\begin{equation}
\label{2.47} \left( \prod_{j\, =\, 1}^{m} \prod_{\mu \, =\, 0}^{3}
\left( \frac{\partial}{\partial x_{j}^{\mu}} \right)^{k_{j}^{\mu}}
\right) \phi (x_{1},...,x_{m})\bigr|_{x_{j}\, =\, 0} =
b(k_{j}^{\mu}).
\end{equation}
It is sufficient to construct a function $\phi (x) \in D({\bf R})$
satisfying one dimensional condition (\ref{2.47})
\begin{equation}
\label{2.48} \left( \frac{d}{dx} \right)^{k} \phi (x)\bigr|_{x\, =\,
0} = b(k).
\end{equation}
Let a function $h(x) \in D({\bf R})$ be equal to $1$ in a
neighborhood of the point $x = 0$. We prove that the function
\begin{equation}
\label{2.49} \phi (x) = \sum_{k\, =\, 0}^{\infty}
a(k)\frac{x^{k}}{k!} h(x)\exp \{ - |a(k)|x^{2} \}
\end{equation}
belongs to the space $D({\bf R})$ for any numbers $a(k)$. It is easy
to obtain the estimates
\begin{equation}
\label{2.50} \biggl| \sum_{k\, =\, 2}^{\infty} a(k)\frac{x^{k}}{k!}
 \exp \{ - |a(k)|x^{2}\} \biggr| \leq \left( \sum_{k\, =\,
2}^{\infty} \frac{|x|^{k - 2}}{k!}\right) \sup_{k\, \in \, {\bf
Z}_{+},\, \, x\, \in \, {\bf R}} |a(k)| x^{2} \exp \{ -
|a(k)|x^{2}\},
\end{equation}
\begin{equation}
\label{2.51} \sup_{k\, \in \, {\bf Z}_{+},\, \, x\, \in \, {\bf R}}
|a(k)| x^{2} \exp \{ - |a(k)|x^{2}\} \leq \sup_{0\, \leq \, x\, <\,
\infty } xe^{- x} = e^{- 1}.
\end{equation}
The estimates (\ref{2.50}), (\ref{2.51}) imply the absolute
convergence of the series (\ref{2.49}). If we differentiate every
term of the series (\ref{2.49}), the obtained series will be
absolutely convergent. The proof is similar to the above one. The
function (\ref{2.49}) is infinitely differentiable and has a compact
support. Hence $\phi (x) \in D({\bf R})$.

The definition (\ref{2.49}) implies
\begin{eqnarray}
\label{2.52} \phi (0) = a(0), \nonumber \\
... \nonumber \\
\left( \frac{d}{dx}\right)^{m} \phi (x)\bigr|_{x\, =\, 0} = a(m) +
\sum_{k\, =\, 0}^{m - 1} \sum_{n\, =\, 0}^{[\frac{m}{2} ]} C_{kn}
a(k)|a(k)|^{n}
\end{eqnarray}
where $C_{kn}$ is a combinatorial coefficient. We solve the
recurrent relations (\ref{2.48}), (\ref{2.52}) and find the
coefficients $a(k)$. The lemma is proved.

The Fourier transform of the distribution (\ref{2.46}) is the power
series. It is natural to define the series (\ref{2.46}) for the
functions $\phi (x_{1},...,x_{m})$ whose Fourier transform
(\ref{2.44}) has a compact support.

\section{PCT, permutation invariance and spin condition}
\setcounter{equation}{0}

We introduce the vacuum expectation values of the product of the
quantum fields where the action of the permutation group is simple.
Let us consider the vectors $p_{ij} = p_{ji} \in {\bf R}^{4}$, $1
\leq i < j \leq n + 1$, $n = 1,2,...$ and the distribution
$$
f(p_{ij}, 1 \leq i < j \leq n + 1) \in S^{\prime}({\bf
R}^{4\frac{1}{2} n(n + 1)})
$$
with support in the product of the cones (\ref{1.7}). The
distribution
\begin{eqnarray}
\label{3.1} \int d^{4n}p \ast f(p_{1},...,p_{n}) \phi
(p_{1},...,p_{n}) =  \int \left(
 \prod_{1\, \leq \, i\, <\, j\, \leq \, n + 1} d^{4}p_{ij}\right) \nonumber \\ f(p_{ij}, 1 \leq i < j \leq n + 1)
\phi \left( \sum_{1\, <\, j\, \leq \, n + 1} p_{1j},...,\sum_{1\,
\leq \, i \, \leq \, k,\, \, k\, <\, j\, \leq \, n + 1}
p_{ij},...,\sum_{1\, \leq \, i\, \leq \, n} p_{i,n + 1}\right)
\end{eqnarray}
is called the multi - convolution of the distribution $f(p_{ij}, 1
\leq i < j \leq n + 1)$.

Let us consider the complex vectors $z_{j} \in {\bf C}^{4}$, $j =
1,...,n + 1$,
$$
\hbox{Im} (z_{j + 1} - z_{j}) \in V_{+},\, \, j = 1,...,n,
$$
$$
V_{+} = \{ x \in {\bf R}^{4}: x^{0} > 0, (x,x) > 0 \}.
$$
Hence
$$
\hbox{Im} (z_{j} - z_{i}) \in V_{+},\, \, 1 \leq i < j \leq n + 1.
$$
It is easy to verify
\begin{equation}
\label{3.2} \sum_{k\, =\, 1}^{n} \left( \sum_{1\, \leq \, i \, \leq
\, k,\, \, k\, <\, j\, \leq \, n + 1} p_{ij},z_{k + 1} -
z_{k}\right) = \sum_{1\, \leq \, i\, <\, j\, \leq \, n + 1}
(p_{ij},z_{j} - z_{i}).
\end{equation}
By making use of the relation (\ref{3.2}) we can calculate the
Fourier - Laplace transform of the multi - convolution (\ref{3.1})
\begin{eqnarray}
\label{3.3} \int d^{4n}p \ast f(p_{1},...,p_{n})\exp \{ \sqrt{- 1}
\sum_{k\, =\, 1}^{n} (p_{k},z_{k + 1} - z_{k})\} = \nonumber \\ \int
\left(
 \prod_{1\, \leq \, i\, <\, j\, \leq \, n + 1} d^{4}p_{ij}\right)
 f(p_{ij}, 1 \leq i < j \leq n + 1)
 \exp \{ \sqrt{- 1} \sum_{1\, \leq \, i\, <\, j\, \leq \, n + 1}
(p_{ij},z_{j} - z_{i})\}.
\end{eqnarray}
The step function
\begin{equation}
\label{3.4} \theta (k) = \left\{ {1, \hskip 0,5cm k = 0,1,2,3,...}
\atop {0, \hskip 0,5cm k = - 1, - 2,...} \right.
\end{equation}
is the restriction of the step function (\ref{2.17}) on the
integers. Let $\pi$ be a permutation of the natural numbers $1,...,n
+ 1$. By using the changes $i \rightarrow \pi (i)$, $j \rightarrow
\pi (j)$ of the summation variables we have
\begin{eqnarray}
\label{3.5} \sum_{1\, \leq \, i\, <\, j\, \leq \, n + 1}
(p_{ij},z_{j} - z_{i}) = \frac{1}{2} \sum_{ i,j\, = \, 1,...,n +
1,\, \, i\, \neq \, j} (p_{ij},(- 1)^{\theta (i - j)}(z_{j} -
z_{i})) = \nonumber \\ \sum_{1\, \leq \, i\, <\, j\, \leq \, n + 1}
(p_{\pi (i)\pi (j)},(- 1)^{\theta (\pi (i) - \pi (j))}(z_{\pi (j)} -
z_{\pi (i)})).
\end{eqnarray}
By making use of the relation (\ref{3.5}) and the change $p_{\pi
(i)\pi (j)} \rightarrow p_{ij}$ of the integration variables we get
\begin{eqnarray}
\label{3.6} \int \left(
 \prod_{1\, \leq \, i\, <\, j\, \leq \, n + 1} d^{4}p_{ij}\right)
 f(p_{\pi (i)\pi (j)}, 1 \leq i < j \leq n + 1) \times \nonumber \\
  \exp \{ \sqrt{- 1} \sum_{1\, \leq \, i\, <\, j\, \leq \, n + 1}
(p_{ij},z_{j} - z_{i})\} = \nonumber \\ \int \left(
 \prod_{1\, \leq \, i\, <\, j\, \leq \, n + 1} d^{4}p_{ij}\right)
 f(p_{ij}, 1 \leq i < j \leq n + 1) \times \nonumber \\
 \exp \{ \sqrt{- 1} \sum_{1\, \leq \, i\, <\, j\, \leq \, n + 1}
(p_{ij},(- 1)^{\theta (\pi (i) - \pi (j))}(z_{\pi (j)} - z_{\pi
(i)}))\}.
\end{eqnarray}
The right - hand side of the equality (\ref{3.6}) and the action of
the permutation $\pi$ on the function (\ref{3.3}) of the variables
$z_{1},...,z_{n + 1}$ differ from each other in the sign multipliers
$(- 1)^{\theta (\pi (i) - \pi (j))}$ only.

Let us consider the distributions (\ref{1.4}) for $n = 1$. The
distribution (\ref{1.6}) has the form (\ref{2.31}), (\ref{2.35}) for
$m = n = 1$. Let the half - integers $l_{2} = \dot{l}_{1}$,
$\dot{l}_{2} = l_{1}$, $l_{3} = l_{12}$, $m_{3} = m_{12}$,
$\dot{m}_{3} = \dot{m}_{12}$. This distribution is the vacuum
expectation value of the product of the quantum field operator and
its adjoint operator. The Clebsch - Gordan coefficient
$C(l_{1},\dot{l}_{1}, l_{12};m_{1},m_{2},m_{12})$ is not zero only
in the case when the half - integer $l_{12}$ is equal to one of the
half - integers $|l_{1} - \dot{l}_{1}|,|l_{1} - \dot{l}_{1}| +
1,....,l_{1} + \dot{l}_{1} - 1,l_{1} + \dot{l}_{1}$. Hence
\begin{equation}
\label{3.7} 2l_{12} = 2l_{1} + 2\dot{l}_{1} = (2l_{1} +
2\dot{l}_{1})^{2} = (2l_{1} + 2\dot{l}_{1})(2l_{2} + 2\dot{l}_{2})\,
\, \, \hbox{mod} \, 2.
\end{equation}
{\bf Spin condition}

\noindent {\it We consider the infinite number of the quantum
fields. Let the set of the natural numbers} $1,...,n + 1$ {\it be
divided into the sets} $A_{1},...,A_{N}$. {\it Some of the sets}
$A_{1},...,A_{N}$ {\it may be empty. With any natural number} $i =
1,...,n + 1$ {\it there correspond the half - integers}
$l_{i},\dot{l}_{i} \in 1/2{\bf Z}_{+}$; $m_{i} = - l_{i}, - l_{i} +
1,...,l_{i} - 1,l_{i}$; $\dot{m}_{i} = - \dot{l}_{i}, - \dot{l}_{i}
+ 1,....,\dot{l}_{i} - 1,\dot{l}_{i}$ {\it such that} $l_{i} =
l_{j}$, $\dot{l}_{i} = \dot{l}_{j}$ {\it if the natural numbers}
$i,j$ {\it belong to the same set} $A_{k}$.

{\it The vacuum expectation value of the product of} $n + 1$, $n =
1,2,...$, {\it quantum fields is the distribution}
\begin{eqnarray}
\label{3.8} W_{m_{1},...,m_{n + 1};\dot{m}_{1},...,\dot{m}_{n +
1}}^{l_{1},...,l_{n + 1};\dot{l}_{1},...,\dot{l}_{n + 1}}
(A_{1},...,A_{N};x_{2} - x_{1},...,x_{n + 1} - x_{n}) = \nonumber
\\ \int \left( \prod_{1\, \leq \, i\, <\, j\, \leq \, n + 1} d^{4}p_{ij}
\right) F_{m_{1},...,m_{n + 1};\dot{m}_{1},...,\dot{m}_{n +
1}}^{l_{1},...,l_{n + 1};\dot{l}_{1},...,\dot{l}_{n + 1}}
(A_{1},...,A_{N};p_{ij}, 1 \leq i < j \leq n + 1)\times \nonumber \\
\exp \{ \sqrt{- 1} \sum_{k\, =\, 1}^{n} \left( \sum_{1\, \leq \, i
\, \leq \, k,\, \, k\, <\, j\, \leq \, n + 1} p_{ij},x_{k + 1} -
x_{k}\right) \}.
\end{eqnarray}
{\it The distribution}
\begin{eqnarray}
\label{3.9} \int \left( \prod_{1\, \leq \, i\, <\, j\, \leq \, n +
1} d^{4}p_{ij} \right) F_{m_{1},...,m_{n +
1};\dot{m}_{1},...,\dot{m}_{n + 1}}^{l_{1},...,l_{n +
1};\dot{l}_{1},...,\dot{l}_{n + 1}}
(A_{1},...,A_{N};p_{ij}, 1 \leq i < j \leq n + 1)\times \nonumber \\
\phi (p_{ij}, 1 \leq i < j \leq n + 1) = \nonumber \\ \sum_{l_{ij}
\in 1/2{\bf Z}_{+},\, \, 1\, \leq \, i\, <\, j\, \leq \, n + 1}
\sum_{m_{ij},\dot{m}_{ij} = - l_{ij}, - l_{ij} + 1,...,l_{ij} -
1,l_{ij},\, \, 1\, \leq \, i\, <\, j\, \leq \, n + 1} \int \left(
\prod_{1\, \leq \, i\, <\, j\, \leq \, n + 1} d^{4}p_{ij} \right)
\nonumber \\ f_{m_{i},\dot{m}_{i},\, \, 1\, \leq \, i\, \leq \, n +
1 ;\, \, m_{ij},\dot{m}_{ij},\, \, 1\, \leq \, i\, <\, j\, \leq \, n
+ 1}^{l_{i},\dot{l}_{i},\, \, 1\, \leq \, i\, \leq \, n + 1;\, \,
l_{ij}\, \, 1\, \leq \, i\, <\, j\, \leq \, n + 1}
(A_{1},...,A_{N};(p_{ij},p_{ij}), 1 \leq i < j \leq n + 1)\times \nonumber \\
\left( \prod_{1\, \leq \, i\, <\, j\, \leq \, n + 1} \theta
(p_{ij}^{0}) (\partial_{p_{ij}}, \partial_{p_{ij}})^{q}
t_{m_{ij}\dot{m}_{ij}}^{l_{ij}} (- \tilde{\partial}_{p_{ij}})
\right) \phi (p_{ij}, 1 \leq i < j \leq n + 1)
\end{eqnarray}
{\it is defined on any test function} $\phi (p_{ij}, 1 \leq i < j
\leq n + 1)$ {\it whose Fourier transform is an infinitely
differentiable function with a compact support.} $q$ {\it is a
natural number.} $2\times 2$ - {\it matrix} $\tilde{\partial}_{x}$
{\it is given by the relation} (\ref{2.19}). {\it The polynomial}
$t_{mn}^{l}(A)$ {\it is given by the relation} (\ref{2.4}). {\it The
continuous function}
\begin{equation}
\label{3.10} f_{m_{i},\dot{m}_{i},\, \, 1\, \leq \, i\, \leq \, n +
1 ;\, \, m_{ij},\dot{m}_{ij},\, \, 1\, \leq \, i\, <\, j\, \leq \, n
+ 1}^{l_{i},\dot{l}_{i},\, \, 1\, \leq \, i\, \leq \, n + 1;\, \,
l_{ij}\, \, 1\, \leq \, i\, <\, j\, \leq \, n + 1} (A_{1},...,A_{N};
s_{ij}, 1 \leq i < j \leq n + 1)
\end{equation}
{\it with a support in the product of the positive semi - axes is
polynomial bounded. The parity of any integer} $2l_{ij}$ {\it in the
equality} (\ref{3.9}) {\it is given by the relation}
\begin{equation}
\label{3.11} 2l_{ij} = (2l_{i} + 2\dot{l}_{i})(2l_{j} +
2\dot{l}_{j})\, \, \hbox{mod} \, 2,\, \, 1\, \leq \, i\, <\, j\,
\leq \, n + 1.
\end{equation}

Any term in the series (\ref{3.9}) is a tempered distribution from
the space $S^{\prime} ({\bf R}^{4\frac{1}{2} n(n + 1)})$. This
distribution has a support in the product of the closed upper light
cones (\ref{1.7}). The support of the infinite series (\ref{3.9}) is
not defined.

\noindent {\bf Theorem 3.1.} {\it Let for any permutation} $\pi$
{\it of the natural numbers} $1,...,n + 1$ {\it the distribution}
(\ref{3.9}) {\it satisfy the relation}
\begin{eqnarray}
\label{3.12} F_{m_{\pi (i)},\dot{m}_{\pi (i)},\, \, 1\, \leq \, i\,
\leq \, n + 1}^{l_{\pi (i)},\dot{l}_{\pi (i)},\, \, 1\, \leq \, i\,
\leq \, n + 1} (\pi (A_{i}), 1 \leq i \leq N; p_{\pi (i)\pi (j)}, 1
\leq i < j \leq n + 1) = \nonumber \\ F_{m_{1},...,m_{n +
1};\dot{m}_{1},...,\dot{m}_{n + 1}}^{l_{1},...,l_{n +
1};\dot{l}_{1},...,\dot{l}_{n + 1}} (A_{1},...,A_{N};p_{ij}, 1 \leq
i < j \leq n + 1)
\end{eqnarray}
{\it Then for the vectors} $x_{i} \in {\bf R}^{4}$, $i = 1,...,n +
1$, $(x_{j} - x_{i},x_{j} - x_{i}) < 0$, $1 \leq i < j \leq n + 1$,
{\it and for any permutation} $\pi$ {\it of the natural numbers}
$1,...,n + 1$ {\it the distribution} (\ref{3.8}) {\it satisfies the
relation}
\begin{eqnarray}
\label{3.13} (- 1)^{\sigma_{l} (\pi)} W_{m_{\pi (i)},\dot{m}_{\pi
(i)},\, \, 1\, \leq \, i\, \leq \, n + 1}^{l_{\pi (i)},\dot{l}_{\pi
(i)},\, \, 1\, \leq \, i\, \leq \, n + 1} (\pi (A_{i}), 1 \leq i
\leq N; x_{\pi (i + 1)} - x_{\pi (i)}, 1 \leq i \leq n) = \nonumber
\\ W_{m_{1},...,m_{n + 1};\dot{m}_{1},...,\dot{m}_{n +
1}}^{l_{1},...,l_{n + 1};\dot{l}_{1},...,\dot{l}_{n + 1}}
(A_{1},...,A_{N};x_{2} - x_{1},...,x_{n + 1} - x_{n})
\end{eqnarray}
{\it where the number}
\begin{equation}
\label{3.14} \sigma_{l} (\pi) = \sum_{1\, \leq \, i\, <\, j\, \leq
\, n + 1} (2l_{i} + 2\dot{l}_{i})(2l_{j} + 2\dot{l}_{j})\theta (\pi
(i) - \pi (j))\, \, \hbox{mod} \, 2.
\end{equation}
{\it Proof.} Let a function $\phi (x_{1},...,x_{n + 1}) \in D({\bf
R}^{4(n + 1)})$. Due to the definitions (\ref{3.8}), (\ref{3.9}) and
the relation (\ref{3.2})
\begin{eqnarray}
\label{3.15} \int d^{4(n + 1)}xW_{m_{1},...,m_{n +
1};\dot{m}_{1},...,\dot{m}_{n + 1}}^{l_{1},...,l_{n +
1};\dot{l}_{1},...,\dot{l}_{n + 1}} (A_{1},...,A_{N};x_{2} -
x_{1},...,x_{n + 1} - x_{n})\phi (x_{1},...,x_{n + 1}) = \nonumber
\\ \sum_{l_{ij}
\in 1/2{\bf Z}_{+},\, \, 1\, \leq \, i\, <\, j\, \leq \, n + 1}
\sum_{m_{ij},\dot{m}_{ij} = - l_{ij}, - l_{ij} + 1,...,l_{ij} -
1,l_{ij},\, \, 1\, \leq \, i\, <\, j\, \leq \, n + 1} \int \left(
\prod_{1\, \leq \, i\, <\, j\, \leq \, n + 1} d^{4}p_{ij} \right)
\nonumber \\ f_{m_{i},\dot{m}_{i},\, \, 1\, \leq \, i\, \leq \, n +
1 ;\, \, m_{ij},\dot{m}_{ij},\, \, 1\, \leq \, i\, <\, j\, \leq \, n
+ 1}^{l_{i},\dot{l}_{i},\, \, 1\, \leq \, i\, \leq \, n + 1;\, \,
l_{ij}\, \, 1\, \leq \, i\, <\, j\, \leq \, n + 1}
(A_{1},...,A_{N};(p_{ij},p_{ij}), 1 \leq i < j \leq n + 1)\times \nonumber \\
\left( \prod_{1\, \leq \, i\, <\, j\, \leq \, n + 1} \theta
(p_{ij}^{0}) (\partial_{p_{ij}}, \partial_{p_{ij}})^{q}
t_{m_{ij}\dot{m}_{ij}}^{l_{ij}} (- \tilde{\partial}_{p_{ij}})
\right) \times \nonumber \\ \int d^{4(n + 1)}x\phi (x_{1},...,x_{n +
1})\exp \{ \sqrt{- 1} \sum_{1\, \leq \, i\, <\, j\, \leq \, n + 1}
(p_{ij},x_{j} - x_{i})\}.
\end{eqnarray}
In view of (\cite{9}, Section 26.3) we have
\begin{eqnarray}
\label{3.16} \int \left( \prod_{1\, \leq \, i\, <\, j\, \leq \, n +
1} d^{4}p_{ij} \right) \nonumber \\ f_{m_{i},\dot{m}_{i},\, \, 1\,
\leq \, i\, \leq \, n + 1 ;\, \, m_{ij},\dot{m}_{ij},\, \, 1\, \leq
\, i\, <\, j\, \leq \, n + 1}^{l_{i},\dot{l}_{i},\, \, 1\, \leq \,
i\, \leq \, n + 1;\, \, l_{ij}\, \, 1\, \leq \, i\, <\, j\, \leq \,
n + 1}
(A_{1},...,A_{N};(p_{ij},p_{ij}), 1 \leq i < j \leq n + 1)\times \nonumber \\
\left( \prod_{1\, \leq \, i\, <\, j\, \leq \, n + 1} \theta
(p_{ij}^{0}) (\partial_{p_{ij}}, \partial_{p_{ij}})^{q}
t_{m_{ij}\dot{m}_{ij}}^{l_{ij}} (- \tilde{\partial}_{p_{ij}})
\right) \times \nonumber \\ \int d^{4(n + 1)}x\phi (x_{1},...,x_{n +
1})\exp \{ \sqrt{- 1} \sum_{1\, \leq \, i\, <\, j\, \leq \, n + 1}
(p_{ij},x_{j} - x_{i})\} = \nonumber \\ \lim_{\hbox{Im} z_{i}\,
\rightarrow \, 0,\, \, \hbox{Im} (z_{j} - z_{i})\, \in \, V_{+},\,
\, 1\, \leq \, i\, <\, j\, \leq \, n + 1} \int \left( \prod_{1\,
\leq \, i\, <\, j\, \leq \, n + 1} d^{4}p_{ij} \right)
 \nonumber \\ f_{m_{i},\dot{m}_{i},\, \, 1\, \leq \, i\, \leq
\, n + 1 ;\, \, m_{ij},\dot{m}_{ij},\, \, 1\, \leq \, i\, <\, j\,
\leq \, n + 1}^{l_{i},\dot{l}_{i},\, \, 1\, \leq \, i\, \leq \, n +
1;\, \, l_{ij}\, \, 1\, \leq \, i\, <\, j\, \leq \, n + 1}
(A_{1},...,A_{N};(p_{ij},p_{ij}), 1 \leq i < j \leq n + 1)\times \nonumber \\
\left( \prod_{1\, \leq \, i\, <\, j\, \leq \, n + 1} \theta
(p_{ij}^{0}) (\partial_{p_{ij}}, \partial_{p_{ij}})^{q}
t_{m_{ij}\dot{m}_{ij}}^{l_{ij}} (- \tilde{\partial}_{p_{ij}})
\right) \times \nonumber \\ \int d^{4(n + 1)}\hbox{Re} z\phi
(\hbox{Re}
 z_{1},...,\hbox{Re} z_{n + 1})\exp \{ \sqrt{- 1} \sum_{1\, \leq \,
i\, <\, j\, \leq \, n + 1} (p_{ij},z_{j} - z_{i})\}.
\end{eqnarray}

Let a function $f(s)$ with a support in the positive semi - axis be
polynomial bounded. We can represent any vector $y \in V_{+}$ in the
form
$$
\tilde{y} = (y,y)^{1/2} g(t,z)g(t,z)^{\ast},
$$
$$
g(t,z) = \left(
\begin{array}{cc}

t^{- 1} & 0 \\

      z & t

\end{array} \right)
$$
where $t$ is a positive number and $z$ is a complex number.

By making the change of the variables
\begin{equation}
\label{3.17} \tilde{p} = \mu^{1/2} g(t,z)u(\phi, \theta, \psi )
((\cosh s)\sigma^{0} + (\sinh s)\sigma^{3})u(\phi, \theta, \psi
)^{\ast} g(t,z)^{\ast}
\end{equation}
where $2\times 2$ - matrix $u(\phi, \theta, \psi )$ is given by the
relation (\ref{2.41}) we have
\begin{eqnarray}
\label{3.18} \int d^{4}p \theta (p^{0})f((p,p))\exp \{ - (p,y)\} =
\nonumber \\ 2\pi \int_{0}^{\infty} d\mu \int_{0}^{\infty} ds
\int_{SU(2)} du(\phi, \theta, \psi ) \mu f(\mu)\sinh^{2} s\exp \{ -
\mu^{1/2} (y,y)^{1/2}\cosh s\}.
\end{eqnarray}
The normalized Haar measure $du(\phi, \theta, \psi )$ on the group
$SU(2)$ is given by the relation (\ref{2.42}). By making use of the
relation (\cite{10}, relation 7.12(21))
\begin{equation}
\label{3.19} K_{\nu}(z) = \int_{0}^{\infty} ds (\cosh \nu s)\exp \{
 - z\cosh s\},\, \, \hbox{Re} z\, >\, 0,
\end{equation}
and the relation (\cite{10}, relation 7.11(25))
\begin{equation}
\label{3.20} K_{\nu + 1}(z) - K_{\nu - 1} = 2\nu z^{- 1}K_{\nu}(z)
\end{equation}
we can rewrite the equality (\ref{3.18}) in the form
\begin{equation}
\label{3.21} \int d^{4}p \theta (p^{0})f((p,p))\exp \{ - (p,y)\} =
2\pi \int_{0}^{\infty} d\mu \mu f(\mu)(\mu (y,y))^{- 1/2}K_{1}((\mu
(y,y))^{1/2}).
\end{equation}
Due to (\cite{10}, relation 7.2.5(37))
\begin{eqnarray}
\label{3.22} \mu (\mu (y,y))^{- 1/2}K_{1}((\mu (y,y))^{1/2}) =
\nonumber \\ \frac{1}{2} \mu \ln \left( \frac{1}{2}(\mu
(y.y))^{1/2}\right) \sum_{m\, =\, 0}^{\infty} \frac{1}{m!(m + 1)!}
\left( \frac{1}{4} \mu (y,y)\right)^{m} + \nonumber \\
 (y,y)^{- 1} - 2^{- 3/2}\mu \sum_{m\, =\, 0}^{\infty} \left( \frac{1}{4} \mu
 (y,y)\right)^{m} \frac{\psi (m + 2) + \psi (m + 1)}{m!(m + 1)!}
\end{eqnarray}
where the logarithmic derivative of the gamma - function
$$
\psi (z) = \frac{d}{dz} (\ln \Gamma (z)).
$$
For $- \frac{3\pi}{2} < \arg z < \frac{3\pi}{2}$ the asymptotic
relation (\cite{10}, relation 7.13(7)) is valid
\begin{equation}
\label{3.23} K_{\nu}(z) = \left( \frac{\pi}{2z}\right)^{1/2} e^{-
z}(1 + O(|z|^{- 1})).
\end{equation}
In view of the relations (\ref{3.22}), (\ref{3.23}) both parts of
the equality (\ref{3.21}) as the functions of the variable $iy$ have
an analytic continuation into the tube domain ${\bf R}^{4} + iV_{+}$
\begin{equation}
\label{3.24} \int d^{4}p \theta (p^{0})f((p,p))\exp \{ i(p,z)\} =
2\pi \int_{0}^{\infty} d\mu \mu f(\mu)(- \mu (z,z))^{- 1/2}K_{1}((-
\mu (z,z))^{1/2})
\end{equation}
where a complex vector $z = x + iy \in {\bf R}^{4} + iV_{+}$.

Let the compact support of the function $\phi (x_{1},...,x_{n + 1})$
lie in the domain $(x_{j} - x_{i},x_{j} - x_{i}) < 0$, $1 \leq i < j
\leq n + 1$. The relations (\ref{3.16}), (\ref{3.22}) - (\ref{3.24})
imply
\begin{eqnarray}
\label{3.25} \int \left( \prod_{1\, \leq \, i\, <\, j\, \leq \, n +
1} d^{4}p_{ij} \right) \nonumber \\ f_{m_{i},\dot{m}_{i},\, \, 1\,
\leq \, i\, \leq \, n + 1 ;\, \, m_{ij},\dot{m}_{ij},\, \, 1\, \leq
\, i\, <\, j\, \leq \, n + 1}^{l_{i},\dot{l}_{i},\, \, 1\, \leq \,
i\, \leq \, n + 1;\, \, l_{ij}\, \, 1\, \leq \, i\, <\, j\, \leq \,
n + 1}
(A_{1},...,A_{N};(p_{ij},p_{ij}), 1 \leq i < j \leq n + 1)\times \nonumber \\
\left( \prod_{1\, \leq \, i\, <\, j\, \leq \, n + 1} \theta
(p_{ij}^{0}) (\partial_{p_{ij}}, \partial_{p_{ij}})^{q}
t_{m_{ij}\dot{m}_{ij}}^{l_{ij}} (- \tilde{\partial}_{p_{ij}})
\right) \times \nonumber \\ \int d^{4(n + 1)}x\phi (x_{1},...,x_{n +
1})\exp \{ \sqrt{- 1} \sum_{1\, \leq \, i\, <\, j\, \leq \, n + 1}
(p_{ij},x_{j} - x_{i})\} = \nonumber \\ \int d^{4(n + 1)}x\phi
(x_{1},...,x_{n + 1}) \times \nonumber \\ \left( \prod_{1\, \leq \,
i\, <\, j\, \leq \, n + 1} (2\pi ( - (x_{j} - x_{i},x_{j} -
x_{i}))^{q}t_{m_{ij}\dot{m}_{ij}}^{l_{ij}}(\sqrt{- 1} (\tilde{x}_{i}
- \tilde{x}_{j})))\right) \times \nonumber
\\ \left( \prod_{1\, \leq \, i\, <\, j\, \leq \, n + 1}
\int_{0}^{\infty} \mu_{ij} d\mu_{ij} \right) \nonumber \\
f_{m_{i},\dot{m}_{i},\, \, 1\, \leq \, i\, \leq \, n + 1 ;\, \,
m_{ij},\dot{m}_{ij},\, \, 1\, \leq \, i\, <\, j\, \leq \, n +
1}^{l_{i},\dot{l}_{i},\, \, 1\, \leq \, i\, \leq \, n + 1;\, \,
l_{ij}\, \, 1\, \leq \, i\, <\, j\, \leq \, n + 1}
(A_{1},...,A_{N};\mu_{ij}, 1 \leq i < j \leq n + 1)\times \nonumber \\
\left( \prod_{1\, \leq \, i\, <\, j\, \leq \, n + 1} ( - \mu_{ij}
(x_{j} - x_{i},x_{j} - x_{i}))^{- 1/2}K_{1}(( - \mu_{ij} (x_{j} -
x_{i},x_{j} - x_{i}))^{1/2})\right)
\end{eqnarray}
We insert the left - hand side of the equality (\ref{3.12}) into the
right - hand side of the equality (\ref{3.8}). In view of the
relations (\ref{3.6}), (\ref{3.25}) the obtained distribution
coincides with the left - hand side of the equality (\ref{3.13}) for
$(x_{j} - x_{i},x_{j} - x_{i}) < 0$, $1 \leq i < j \leq n + 1$. (The
polynomial (\ref{2.4}) is homogeneous of the degree $2l$. The parity
of the integer $2l_{ij}$ is given by the relation (\ref{3.11}).) The
theorem is proved.

The quantum field of the number $i$ is called anti - commuting if
\begin{equation}
\label{3.26} 2l_{i} + 2\dot{l}_{i} = 1 \, \, \hbox{mod} \, 2.
\end{equation}
The definition (\ref{3.26}) is consistent with the relation
(\ref{3.7}). The set of numbers corresponding to the anti -
commuting quantum fields is denoted by $F$. The definitions
(\ref{3.14}), (\ref{3.26}) imply
\begin{equation}
\label{3.27} \sigma_{l} (\pi) = \sum_{i,j\, \in \, F,\, \, i \neq j}
\theta (\pi (i) - \pi (j))\theta (j - i)\, \, \hbox{mod} \, 2.
\end{equation}
According to the relation (\ref{3.27}) the multiplier $(-
1)^{\sigma_{l} (\pi)}$ in the relation (\ref{3.13}) coincides with
the multiplier $(- 1)^{M}$ in the relation (\ref{1.12}).

\noindent {\bf Lemma 3.2.} {\it Let the permutation} $\lambda$ {\it
of the natural numbers} $1,...,n + 1$ {\it transform the set} $F$
{\it into itself. Then for any permutation} $\pi$ {\it of the
natural numbers} $1,...,n + 1$ {\it the following relation is valid}
\begin{equation}
\label{3.28} \sigma_{l} (\pi \circ \lambda) = \sigma_{l} (\pi) +
\sigma_{l} (\lambda) \, \, \hbox{mod} \, 2.
\end{equation}
{\it Proof.} For any natural numbers $i,j \in F$, $i \neq j$, the
definition (\ref{3.4}) implies
$$
\theta (\lambda (i) - \lambda (j)) + \theta (\lambda (j) - \lambda
(i)) = 1.
$$
By using this relation and the relation (\ref{3.27}) we get
\begin{eqnarray}
\label{3.29} \sigma_{l} (\pi \circ \lambda) = \sum_{i,j\, \in \,
F,\, \, i\, \neq \, j} \theta (\pi (\lambda (i)) - \pi (\lambda
(j)))\theta (\lambda (j) - \lambda (i))\theta (j - i) + \nonumber \\
\sum_{i,j\, \in \, F,\, \, i\, \neq \, j} \theta (\pi (\lambda (i))
- \pi (\lambda (j)))\theta (\lambda (i) - \lambda (j))\theta (j -
i)\, \, \hbox{mod} \, 2.
\end{eqnarray}
For any natural numbers $i,j \in F$, $i \neq j$,
$$
\theta (j - i) = 1 + \theta (i - j)\, \, \hbox{mod} \, 2.
$$
The substitution of this relation into the right - hand side of the
equality (\ref{3.29}) yields
\begin{eqnarray}
\label{3.30} \sigma_{l} (\pi \circ \lambda) = \sum_{i,j\, \in \,
F,\, \, i\, \neq \, j} \theta (\pi (\lambda (i)) - \pi (\lambda
(j)))\theta (\lambda (j) - \lambda (i)) + \nonumber \\ \sum_{i,j\,
\in \, F,\, \, i\, \neq \, j} \theta (\pi (\lambda (i)) - \pi
(\lambda (j)))\theta (\lambda (j) - \lambda (i))\theta (i - j) +
\nonumber \\ \sum_{i,j\, \in \, F,\, \, i\, \neq \, j} \theta (\pi
(\lambda (i)) - \pi (\lambda (j)))\theta (\lambda (i) - \lambda (j))
\theta (j - i)\, \, \hbox{mod} \, 2.
\end{eqnarray}
The permutation $\lambda$ transforms the set $F$ into itself. Let
$\lambda^{- 1}$ be the inverse permutation. We change the summation
variables $i \rightarrow \lambda^{- 1} (i)$, $j \rightarrow
\lambda^{- 1} (j)$ in the first sum in the right - hand side of the
equality (\ref{3.30}) and $i \rightarrow j$, $j \rightarrow i$ in
the second sum. Now the relation (\ref{3.30}) and the relation
$$
\theta (\pi (\lambda (i)) - \pi (\lambda (j))) + \theta (\pi
(\lambda (j)) - \pi (\lambda (i))) = 1
$$
imply the equality (\ref{3.28}). The lemma is proved.

For the permutation $\tau (i) = n + 2 - i$, $i = 1,...,n + 1$, we
have
\begin{eqnarray}
\label{3.31} \sum_{1\, \leq \, i\, <\, j\, \leq \, n + 1} (p_{ij},(-
1)^{\theta (\tau (i) - \tau (j))}(z_{\tau (j)} - z_{\tau (i)})) =
\nonumber \\ \sum_{k\, =\, 1}^{n} \left( \sum_{1\, \leq \, i\, \leq
\, k,\, \, k\, <\, j\, \leq \, n + 1} p_{ij},z_{n + 2 - k} - z_{n +
1 - k}\right).
\end{eqnarray}
Suppose that the series (\ref{3.9}) are finite. Then the equalities
(\ref{3.8}), (\ref{3.31}) and the relations (\ref{3.6}),
(\ref{3.12}) for the permutation $\tau$ imply
\begin{eqnarray}
\label{3.32} W_{m_{1},...,m_{n + 1};\dot{m}_{1},...,\dot{m}_{n +
1}}^{l_{1},...,l_{n + 1};\dot{l}_{1},...,\dot{l}_{n + 1}}
(A_{1},...,A_{N};x_{2} - x_{1},...,x_{n + 1} - x_{n}) = \nonumber
\\ W_{m_{n + 1},...,m_{1};\dot{m}_{n + 1},...,\dot{m}_{
1}}^{l_{n + 1},...,l_{1};\dot{l}_{n + 1},...,\dot{l}_{1}} (\tau
(A_{1}),...,\tau (A_{N});x_{n + 1} - x_{n},x_{n} - x_{n -
1},...,x_{2} - x_{1}).
\end{eqnarray}
In order to obtain the equality (\ref{3.32}) we use also the fact
that the distribution (\ref{3.8}) is a boundary value of the Fourier
- Laplace transform of the multi - convolution  (\ref{3.1}) of the
distribution (\ref{3.9}) (\cite{9}, Section 26.3).

Due to (\cite{2}, relation (4 - 30)) the vacuum expectation value of
the product of $n + 1$ quantum fields is the distribution
\begin{equation}
\label{3.33} W_{\mu \cdots \nu}(\xi_{1},...,\xi_{n}) = \left(
\Psi_{0},\psi_{\mu} (x_{1}) \cdots \psi_{\nu} (x_{n + 1})\Psi_{0}
\right)
\end{equation}
where $\xi_{j} = x_{j} - x_{j + 1}$, $j = 1,...,n$. In view of
(\cite{2}, Theorem 4 - 7) the condition of the invariance under PCT
transformation is
\begin{equation}
\label{3.34} \left( \Psi_{0},\psi_{\mu} (x_{1}) \cdots \psi_{\nu}
(x_{n + 1})\Psi_{0} \right) = i^{F}(- 1)^{J}\left(
\Psi_{0},\psi_{\nu} (- x_{n + 1}) \cdots \psi_{\mu} (- x_{ 1})
\Psi_{0} \right).
\end{equation}
The multipliers $(- 1)^{J}$ and $i^{F}$ are defined by the following
relations (\cite{2}, relations (4 - 32) and (4 - 36))
\begin{equation}
\label{3.35} W_{\mu \cdots \nu}(\xi_{1},...,\xi_{n}) = (-
1)^{J}W_{\mu \cdots \nu}(- \xi_{1},...,- \xi_{n}),
\end{equation}
\begin{equation}
\label{3.36} \left( \Psi_{0},\psi_{\mu} (x_{1}) \cdots \psi_{\nu}
(x_{n + 1})\Psi_{0} \right) = i^{F}\left( \Psi_{0},\psi_{\nu} (x_{n
+ 1}) \cdots \psi_{\mu} (x_{ 1}) \Psi_{0} \right).
\end{equation}
The function (\ref{3.33}) is holomorphic at the point
$x_{1},...,x_{n + 1}$.

Let us prove that for the finite series (\ref{3.9}) the distribution
(\ref{3.8}), (\ref{3.9}) satisfies the relations (\ref{3.35}),
(\ref{3.36}) and the multiplier $i^{F}(- 1)^{J} = 1$. Therefore PCT
invariance corresponds with the relation (\ref{3.12}) for the
permutation $\tau (i) = n + 2 - i$, $i = 1,...,n + 1$.

In view of the equalities (\ref{3.23}), (\ref{3.25}) for the finite
series (\ref{3.9}) the function (\ref{3.8}), (\ref{3.9}) is
holomorphic at the point $(x_{j} - x_{i},x_{j} - x_{i}) < 0$, $1
\leq i < j \leq n + 1$. Then the relation (\ref{3.13}) for the
permutation $\tau$ corresponds to the relation (\ref{3.36}). The
multiplier
\begin{equation}
\label{3.37} i^{F} = (- 1)^{\sigma_{l} (\tau)} = (- 1)^{\sum_{1\,
\leq \, i\, <\, j\, \leq \, n + 1} (2l_{i} + 2\dot{l}_{i})(2l_{j} +
2\dot{l}_{j})}.
\end{equation}
The equalities (\ref{3.11}), (\ref{3.25}) imply
\begin{eqnarray}
\label{3.38} W_{m_{1},...,m_{n + 1};\dot{m}_{1},...,\dot{m}_{n +
1}}^{l_{1},...,l_{n + 1};\dot{l}_{1},...,\dot{l}_{n + 1}}
(A_{1},...,A_{N};x_{2} - x_{1},...,x_{n + 1} - x_{n}) = \nonumber \\
(- 1)^{J}W_{m_{1},...,m_{n + 1};\dot{m}_{1},...,\dot{m}_{n +
1}}^{l_{1},...,l_{n + 1};\dot{l}_{1},...,\dot{l}_{n + 1}}
(A_{1},...,A_{N};- (x_{2} - x_{1}),...,- (x_{n + 1} - x_{n})),
\end{eqnarray}
\begin{equation}
\label{3.39} (- 1)^{J} = (- 1)^{\sum_{1\, \leq \, i\, <\, j\, \leq
\, n + 1} (2l_{i} + 2\dot{l}_{i})(2l_{j} + 2\dot{l}_{j})}.
\end{equation}
The relation (\ref{3.38}) corresponds to the relation (\ref{3.35}).
The multipliers (\ref{3.37}), (\ref{3.39}) satisfy the relation
$i^{F}(- 1)^{J} = 1$.

\noindent {\bf Permutation invariance}

\noindent {\it The vacuum expectation value of the product of} $n +
1$, $n = 1,2,...$, {\it quantum fields is the distribution}
(\ref{3.8}), (\ref{3.9}) {\it where the functions} (\ref{3.10}) {\it
satisfy the relation}
\begin{eqnarray}
\label{3.40} f_{m_{\pi (i)},\dot{m}_{\pi (i)},\, \, 1\, \leq \, i\,
\leq \, n + 1 ;\, \, m_{\pi (i) \pi (j)},\dot{m}_{\pi (i) \pi
(j)},\, \, 1\, \leq \, i\, <\, j\, \leq \, n + 1}^{l_{\pi
(i)},\dot{l}_{\pi (i)},\, \, 1\, \leq \, i\, \leq \, n + 1;\, \,
l_{\pi (i) \pi (j)}\, \, 1\, \leq \, i\, <\, j\, \leq \, n + 1}
\nonumber \\ (\pi (A_{i}),1 \leq i \leq N; s_{\pi (i) \pi (j)}, 1 \leq i < j \leq n + 1)
 = \nonumber \\
f_{m_{i},\dot{m}_{i},\, \, 1\, \leq \, i\, \leq \, n + 1 ;\, \,
m_{ij},\dot{m}_{ij},\, \, 1\, \leq \, i\, <\, j\, \leq \, n +
1}^{l_{i},\dot{l}_{i},\, \, 1\, \leq \, i\, \leq \, n + 1;\, \,
l_{ij}\, \, 1\, \leq \, i\, <\, j\, \leq \, n + 1} (A_{i},1 \leq i
\leq N; s_{ij}, 1 \leq i < j \leq n + 1)
\end{eqnarray}
{\it for any permutation} $\pi$ {\it of the natural numbers}
$1,...,n + 1$.

\noindent {\bf Lemma 3.3.} {\it The equality} (\ref{3.40}) {\it
implies the equality} (\ref{3.12}).

\noindent {\it Proof.} Since $p_{ji} = p_{ij}$, $l_{ji} = l_{ij}$,
$m_{ji} = m_{ij}$, $\dot{m}_{ji} = \dot{m}_{ij}$ we have for any
permutation $\pi$ of the natural numbers $1,...,n + 1$
\begin{eqnarray}
\label{3.41} \prod_{1\, \leq \, i\, <\, j\, \leq \, n + 1} \theta
(p_{ij}^{0}) (\partial_{p_{ij}}, \partial_{p_{ij}})^{q}
t_{m_{ij}\dot{m}_{ij}}^{l_{ij}} (- \tilde{\partial}_{p_{ij}}) =
\nonumber \\ \prod_{1\, \leq \, i\, <\, j\, \leq \, n + 1} \theta
(p_{\pi (i) \pi (j)}^{0}) (\partial_{p_{\pi (i) \pi
(j)}},\partial_{p_{\pi (i) \pi (j)}})^{q} t_{m_{\pi (i) \pi (j)}
\dot{m}_{\pi (i) \pi (j)}}^{l_{\pi (i) \pi (j)}} (-
\tilde{\partial}_{p_{\pi (i) \pi (j)}}).
\end{eqnarray}
By making use of the relation (\ref{3.41}) and the changes of the
summation variables $l_{\pi (i) \pi (j)} \rightarrow l_{ij}$,
$m_{\pi (i) \pi (j)} \rightarrow m_{ij}$, $\dot{m}_{\pi (i) \pi (j)}
\rightarrow \dot{m}_{ij}$ we get
\begin{eqnarray}
\label{3.42} \sum_{l_{ij} \in 1/2{\bf Z}_{+},\, \, 1\, \leq \, i\,
<\, j\, \leq \, n + 1} \sum_{m_{ij},\dot{m}_{ij} = - l_{ij}, -
l_{ij} + 1,...,l_{ij} - 1,l_{ij},\, \, 1\, \leq \, i\, <\, j\, \leq
\, n + 1} \int \left( \prod_{1\, \leq \, i\, <\, j\, \leq \, n + 1}
d^{4}p_{ij} \right) \nonumber \\ f_{m_{\pi (i)},\dot{m}_{\pi (i)},\,
\, 1\, \leq \, i\, \leq \, n + 1 ;\, \, m_{\pi (i) \pi
(j)},\dot{m}_{\pi (i) \pi (j)},\, \, 1\, \leq \, i\, <\, j\, \leq \,
n + 1}^{l_{\pi (i)},\dot{l}_{\pi (i)},\, \, 1\, \leq \, i\, \leq \,
n + 1;\, \, l_{\pi (i) \pi (j)}\, \, 1\, \leq \, i\, <\, j\, \leq \,
n + 1} \nonumber \\ (\pi (A_{i}),1 \leq i \leq N; (p_{\pi (i) \pi
(j)},p_{\pi (i) \pi (j)}),
1 \leq i < j \leq n + 1)\times \nonumber \\
\left( \prod_{1\, \leq \, i\, <\, j\, \leq \, n + 1} \theta
(p_{ij}^{0}) (\partial_{p_{ij}}, \partial_{p_{ij}})^{q}
t_{m_{ij}\dot{m}_{ij}}^{l_{ij}} (- \tilde{\partial}_{p_{ij}})
\right) \phi (p_{ij}, 1 \leq i < j \leq n + 1) = \nonumber \\ \int
\left( \prod_{1\, \leq \, i\, <\, j\, \leq \, n + 1} d^{4}p_{ij}
\right) \nonumber \\ F_{m_{\pi (i)},\dot{m}_{\pi (i)},1 \leq i \leq
n + 1}^{l_{\pi (i)},\dot{l}_{\pi (i)},1 \leq i \leq n + 1} (\pi
(A_{i}),1 \leq i \leq N; p_{\pi (i) \pi (j)}, 1 \leq i < j \leq n +
1)
\times \nonumber \\
\phi (p_{ij}, 1 \leq i < j \leq n + 1).
\end{eqnarray}
We insert the left - hand side of the equality (\ref{3.40}) into the
right - hand side of the equality (\ref{3.9}) instead of the
function (\ref{3.10}). The obtained distribution coincides with the
left - hand side of the equality (\ref{3.42}). Hence the equalities
(\ref{3.40}), (\ref{3.42}) imply the equality (\ref{3.12}). The
lemma is proved.

\noindent {\bf Lorentz covariance}

\noindent {\it The vacuum expectation value of the product of} $n +
1$, $n = 1,2,...$, {\it quantum fields is the distribution}
(\ref{3.8}), (\ref{3.9}).

{\it For} $n = 1$ {\it the function} (\ref{3.10}) {\it is}
\begin{eqnarray}
\label{3.43} f_{m_{1},m_{2};\dot{m}_{1}, \dot{m}_{2};
m_{12},\dot{m}_{12}}^{l_{1},l_{2}; \dot{l}_{1}, \dot{l}_{2}; l_{12}}
(A_{1},...,A_{N}; s_{12}) = \nonumber \\
f^{l_{1},l_{2}; \dot{l}_{1}, \dot{l}_{2}; l_{12}} (A_{1},...,A_{N};
s_{12})C(l_{1},l_{2},l_{12};m_{1},m_{2},m_{12})C(\dot{l}_{1},
\dot{l}_{2}, l_{12};\dot{m}_{1}, \dot{m}_{2}, \dot{m}_{12})
\end{eqnarray}
{\it where a continuous function} $f^{l_{1},l_{2}; \dot{l}_{1},
\dot{l}_{2}; l_{12}} (A_{1},...,A_{N}; s_{12})$ {\it with a support
in the positive semi - axis is polynomial bounded; the Clebsch -
Gordan coefficient} $C(l_{1},l_{2},l_{3};m_{1},m_{2},m_{3})$ {\it is
given by the relation} (\ref{2.8}).

{\it For} $n > 1$ {\it the function} (\ref{3.10}) {\it is}
\begin{eqnarray}
\label{3.44} f_{m_{i},\dot{m}_{i},\, \, 1\, \leq \, i\, \leq \, n +
1 ;\, \, m_{ij},\dot{m}_{ij},\, \, 1\, \leq \, i\, <\, j\, \leq \, n
+ 1}^{l_{i},\dot{l}_{i},\, \, 1\, \leq \, i\, \leq \, n + 1;\, \,
l_{ij}\, \, 1\, \leq \, i\, <\, j\, \leq \, n + 1} (A_{1},...,A_{N};
s_{ij}, 1 \leq i < j \leq n + 1) = \nonumber \\
\sum_{j_{1},...,j_{\frac{1}{2} n(n + 3) - 2};j_{1}^{\prime},...,
j_{\frac{1}{2} n(n + 3) - 2}^{\prime}\, \in \, 1/2{\bf Z}_{+}}
\nonumber \\ f_{j_{i},j_{i}^{\prime},\, \, 1\, \leq \, i\, \leq \,
\frac{1}{2} n(n + 3) - 2}^{l_{i},\dot{l}_{i},\, \, 1\, \leq \, i\,
\leq \, n + 1;\, \, l_{ij}\, \, 1\, \leq \, i\, <\, j\, \leq \, n +
1} (A_{1},...,A_{N}; s_{ij}, 1 \leq i < j \leq n + 1)\times
\nonumber
\\ C(l_{1},...,l_{n + 1};l_{ij},1 \leq i < j \leq n +
1;j_{1},...,j_{\frac{1}{2} n(n + 3) - 2};\nonumber
\\ m_{1},...,m_{n +
1};m_{ij}, 1 \leq i < j \leq n + 1)\times \nonumber \\
C(\dot{l}_{1},..., \dot{l}_{n + 1};l_{ij},1 \leq i < j \leq n +
1;j_{1}^{\prime},...,j_{\frac{1}{2} n(n + 3) - 2}^{\prime};\nonumber
\\ \dot{m}_{1},...,\dot{m}_{n + 1};\dot{m}_{ij}, 1 \leq i < j \leq n +
1)
\end{eqnarray}
{\it where a continuous function}
$$
f_{j_{i},j_{i}^{\prime},\, \, 1\, \leq \, i\, \leq \, \frac{1}{2}
n(n + 3) - 2}^{l_{i},\dot{l}_{i},\, \, 1\, \leq \, i\, \leq \, n +
1;\, \, l_{ij}\, \, 1\, \leq \, i\, <\, j\, \leq \, n + 1}
(A_{1},...,A_{N}; s_{ij}, 1 \leq i < j \leq n + 1)
$$
{\it with a support in the product of the positive semi - axes is
polynomial bounded; the generalized Clebsch - Gordan coefficient}
$$
C(l_{1},...,l_{n_{1} + 2};l_{n_{1} + 3},...,l_{n_{1} + n_{2} +
4};j_{1},...,j_{n_{1} + n_{2} + 1};
  m_{1},...,m_{n_{1} + 2};m_{n_{1} + 3},...,m_{n_{1} + n_{2} + 4})
$$
{\it is given by the relation} (\ref{2.32}).

The relations (\ref{2.12}), (\ref{2.13}) imply that the distribution
(\ref{3.9}) for $n = 1$ with the functions (\ref{3.43}) satisfies
the covariance relation (\ref{2.14}). The relations (\ref{2.33}),
(\ref{2.34}) imply that the distribution (\ref{3.9}) with the
functions (\ref{3.44}) satisfies  the covariance relation
(\ref{2.14}).

\section{Asymptotic condition}
\setcounter{equation}{0}

It is proved in the book \cite{2} that G\aa rding - Wightman axioms
\cite{1} including the uniqueness of the vacuum imply
\begin{eqnarray}
\label{4.1} \lim_{t \rightarrow \infty} \left( \Psi_{0},
\psi_{\mu_{1}} (x_{1}) \cdots \psi_{\mu_{j}} (x_{j}) \psi_{\mu_{j +
1}} (x_{j + 1} + ta) \psi_{\mu_{j + 2}} (x_{j + 2} + ta)\cdots
 \psi_{\mu_{n}} (x_{n} + ta) \Psi_{0} \right) = \nonumber \\
\left( \Psi_{0}, \psi_{\mu_{1}} (x_{1}) \cdots \psi_{\mu_{j}}
(x_{j})\Psi_{0} \right) \left( \Psi_{0}, \psi_{\mu_{j + 1}} (x_{j +
1})\psi_{\mu_{j + 2}} (x_{j + 2})\cdots \psi_{\mu_{n}} (x_{n})
\Psi_{0} \right)
\end{eqnarray}
for $a \in {\bf R}^{4}$, $(a,a) < 0$. The limit (\ref{4.1}) is
convergent in the topology of the space $S^{\prime}({\bf R}^{4n})$.

Let us consider the asymptotic behavior of the vacuum expectation
values (\ref{3.8}), (\ref{3.9}). Due to (\cite{11}, Section 3.1) we
define the quasi - asymptotic value of a distribution. Let a
distribution $f(x) \in S^{\prime}({\bf R})$ have a support in the
positive semi - axis. A distribution $f(x)$ has a quasi - asymptotic
value with respect to the transformations $x \rightarrow t^{- 1}x$
and the function $\rho (t) = t^{- \lambda}$, $\lambda$ is a complex
number, if in the topology of the space $S^{\prime}({\bf R})$
\begin{equation}
\label{4.2} \lim_{t \rightarrow \infty} t^{\lambda}f(t^{- 1}x) =
g(x).
\end{equation}
The limit (\ref{4.2}) is the characteristic of the asymptotic
behavior of the distribution $f(x)$ at zero point. If we choose the
transformations $x \rightarrow tx$, then such limit will be the
characteristic of the asymptotic behavior of the distribution $f(x)$
at infinity. We can change the function $\rho (t) = t^{- \lambda}$
in the definition (\ref{4.2}) by any regular varying function
(\cite{11}, Section 3.2).

The definition (\ref{4.2}) implies that the distribution $g(x)$ is
homogeneous of the degree $\lambda$
\begin{equation}
\label{4.3} g(tx) = t^{\lambda}g(x),\, \, t > 0.
\end{equation}
For $\hbox{Re} \lambda > - 1$ any homogeneous distribution $g(x) \in
S^{\prime}({\bf R})$ of the degree $\lambda$ with a support in the
positive semi - axis has the form (\cite{12}, Chapter I, Section
3.11)
\begin{equation}
\label{4.4} g(x) = C\frac{\theta (x)x^{\lambda}}{\Gamma (\lambda +
1)}
\end{equation}
where $C$ is a constant, $\theta (x)$ is the step function
(\ref{2.17}) and $\Gamma (\lambda + 1)$ is the gamma - function. Due
to (\cite{12}, Chapter I, Sections 3.2, 3.5) the function
(\ref{4.4}) of the variable $\lambda$ has the analytic continuation
to an entire function. The formula (\ref{4.4}) is valid for any
complex $\lambda$. Due to (\cite{12}, Chapter I, Section 3.5)
\begin{equation}
\label{4.5} \frac{\theta (x)x^{\lambda}}{\Gamma (\lambda +
1)}\Bigr|_{\lambda \, =\, - n} = \delta^{(n - 1)} (x),\, \, n =
1,2,... .
\end{equation}
The vacuum expectation values (\ref{3.8}), (\ref{3.9}) are defined
by the continuous functions (\ref{3.10}). The quasi - asymptotic
value of the continuous function (\ref{3.10}) we define as the limit
of the type (\ref{4.2}) for the space of the continuous test
functions $\phi (x)$ rapidly decreasing: the norm
$$
\sup_{x\, \in \, {\bf R}} (1 + x^{2})^{m}|\phi (x)|
$$
is finite for any natural number $m$.

Let an infinitely differentiable function $\phi (x_{1},...,x_{n +
1})$ have a compact support in the domain $(x_{j} - x_{i},x_{j} -
x_{i}) < 0$, $1 \leq i < j \leq n + 1$. The relations (\ref{3.8}),
(\ref{3.9}), (\ref{3.15}), (\ref{3.25}) imply
\begin{eqnarray}
\label{4.6} \int d^{4(n + 1)}xW_{m_{1},...,m_{n +
1};\dot{m}_{1},...,\dot{m}_{n + 1}}^{l_{1},...,l_{n +
1};\dot{l}_{1},...,\dot{l}_{n + 1}} (A_{1},...,A_{N};x_{2} -
x_{1},...,x_{n + 1} - x_{n})\phi (x_{1},...,x_{n + 1}) = \nonumber
\\ \sum_{l_{ij}
\in 1/2{\bf Z}_{+},\, \, 1\, \leq \, i\, <\, j\, \leq \, n + 1}
\sum_{m_{ij},\dot{m}_{ij} = - l_{ij}, - l_{ij} + 1,...,l_{ij} -
1,l_{ij},\, \, 1\, \leq \, i\, <\, j\, \leq \, n + 1} \int d^{4(n +
1)}x\phi (x_{1},...,x_{n + 1}) \times \nonumber \\ \left( \prod_{1\,
\leq \, i\, <\, j\, \leq \, n + 1} (2\pi ( - (x_{j} - x_{i},x_{j} -
x_{i}))^{q}t_{m_{ij}\dot{m}_{ij}}^{l_{ij}}(\sqrt{- 1} (\tilde{x}_{i}
- \tilde{x}_{j})))\right) \times \nonumber
\\ \left( \prod_{1\, \leq \, i\, <\, j\, \leq \, n + 1}
\int_{0}^{\infty} \mu_{ij} d\mu_{ij} \right) \nonumber \\
f_{m_{i},\dot{m}_{i},\, \, 1\, \leq \, i\, \leq \, n + 1 ;\, \,
m_{ij},\dot{m}_{ij},\, \, 1\, \leq \, i\, <\, j\, \leq \, n +
1}^{l_{i},\dot{l}_{i},\, \, 1\, \leq \, i\, \leq \, n + 1;\, \,
l_{ij}\, \, 1\, \leq \, i\, <\, j\, \leq \, n + 1}
(A_{1},...,A_{N};\mu_{ij}, 1 \leq i < j \leq n + 1)\times \nonumber \\
\prod_{1\, \leq \, i\, <\, j\, \leq \, n + 1} ( - \mu_{ij} (x_{j} -
x_{i},x_{j} - x_{i}))^{- 1/2}K_{1}(( - \mu_{ij} (x_{j} - x_{i},x_{j}
- x_{i}))^{1/2}).
\end{eqnarray}
{\bf Asymptotic condition}

\noindent {\it The vacuum expectation value of the product of} $n +
1$, $n = 1,2,...$, {\it quantum fields is given by the distribution}
(\ref{3.8}), (\ref{3.9}).

{\it For any natural numbers} $n > n_{1}$, {\it any permutation}
$\pi$ {\it of the natural numbers} $1,...,n + 1$ {\it and any
continuous rapidly decreasing function} $\phi (\mu_{ij}, 1 \leq i <
j \leq n + 1)$ {\it the function} (\ref{3.10}) {\it satisfies the
following condition}
\begin{eqnarray}
\label{4.7} \lim_{t \rightarrow \infty}  \left( \prod_{1\, \leq \,
i\, <\, j\, \leq \, n_{1} + 1,\, \, n_{1} + 1\, <\, i\, <\, j\, \leq
\, n + 1} \int_{0}^{\infty} d\mu_{\pi (i) \pi (j)}\right) \times
\nonumber \\ \left( \prod_{1\, \leq \, i\, \leq \, n_{1} + 1,\, \,
n_{1} + 1\, <\, j\, \leq \, n + 1} \int_{0}^{\infty} t^{q\, -\, 2\,
+\, l_{\pi (i) \pi (j)}} d\mu_{\pi (i) \pi (j)}\right)
f_{m_{i},\dot{m}_{i},\, \, 1\, \leq \, i\, \leq \, n + 1 ;\, \,
m_{ij},\dot{m}_{ij},\, \, 1\, \leq \, i\, <\, j\, \leq \, n +
1}^{l_{i},\dot{l}_{i},\, \, 1\, \leq \, i\, \leq \, n + 1;\, \,
l_{ij}\, \, 1\, \leq \, i\, <\, j\, \leq \, n + 1} \nonumber \\
(A_{1},...,A_{N};\mu_{\pi (i) \pi (j)}, 1 \leq i < j \leq n_{1} + 1,
n_{1} + 1 < i < j \leq n + 1; \nonumber \\ t^{- 1}\mu_{\pi (i) \pi
(j)}, 1 \leq i \leq n_{1} + 1 < j \leq n + 1)\times \nonumber \\
\phi (\mu_{ij}, 1 \leq i < j \leq n + 1) = \left( \prod_{1\, \leq \,
i\, <\, j\, \leq \, n + 1} \int_{0}^{\infty} d\mu_{ij} \right)
\nonumber \\ \hbox{as}(f)_{m_{i},\dot{m}_{i},\, \, 1\, \leq \, i\,
\leq \, n + 1 ;\, \, m_{ij},\dot{m}_{ij},\, \, 1\, \leq \, i\, <\,
j\, \leq \, n + 1}^{l_{i},\dot{l}_{i},\, \, 1\, \leq \, i\, \leq \,
n + 1;\, \, l_{ij},\, \, 1\, \leq \, i\, <\, j\, \leq \, n + 1} (\pi
;A_{1},...,A_{N};\mu_{ij}, 1 \leq i < j \leq n + 1)\times \nonumber
\\ \phi (\mu_{ij}, 1 \leq i < j \leq n + 1)
\end{eqnarray}
{\it where a continuous function}
\begin{equation}
\label{4.8} \hbox{as}(f)_{m_{i},\dot{m}_{i},\, \, 1\, \leq \, i\,
\leq \, n + 1 ;\, \, m_{ij},\dot{m}_{ij},\, \, 1\, \leq \, i\, <\,
j\, \leq \, n + 1}^{l_{i},\dot{l}_{i},\, \, 1\, \leq \, i\, \leq \,
n + 1;\, \, l_{ij},\, \, 1\, \leq \, i\, <\, j\, \leq \, n + 1} (\pi
;A_{1},...,A_{N};\mu_{ij}, 1 \leq i < j \leq n + 1)
\end{equation}
{\it is polynomial bounded.}

\noindent {\bf Lemma 4.1.} {\it Let the vacuum expectation value of
the product of} $n + 1$, $n = 1,2,...$, {\it quantum field be the
distribution} (\ref{3.8}), (\ref{3.9}) {\it with the finite series.
Let the continuous functions} (\ref{3.10}) {\it satisfy the
condition} (\ref{4.7}). {\it Let an infinitely differentiable
function} $\phi (x_{\pi (i)}, 1 \leq i \leq n + 1)$ {\it have a
compact support in the domain} $(x_{\pi (j)} - x_{\pi (i)},x_{\pi
(j)} - x_{\pi (i)}) < 0$, $1 \leq i < j \leq n_{1} + 1$, $n_{1} + 1
< i < j \leq n + 1$. {\it Then for any natural numbers} $n > n_{1}$,
{\it any permutation} $\pi$ {\it of the natural numbers} $1,...,n +
1$ {\it and any vector} $a \in {\bf R}^{4}$, $(a,a) < 0$,
\begin{eqnarray}
\label{4.9} \lim_{t \rightarrow \infty} \int d^{4(n + 1)}x
W_{m_{1},...,m_{n + 1}; \dot{m}_{1},..., \dot{m}_{n + 1}}^{l_{1}
,...,l_{n + 1}; \dot{l}_{1},..., \dot{l}_{n + 1}}
(A_{1},...,A_{N};x_{2} - x_{1},...,x_{n + 1} - x_{n}) \times
\nonumber \\ \phi (x_{\pi (i)}, 1 \leq i \leq n_{1} + 1; x_{\pi (i)}
- ta, n_{1} + 1 < i \leq n + 1) = \nonumber \\ \sum_{l_{ij} \in
1/2{\bf Z}_{+}, \, \, 1\,
 \leq \, i\, <\, j\, \leq \, n + 1} \sum_{m_{ij},\dot{m}_{ij} = -
l_{ij}, - l_{ij} + 1,...,l_{ij} - 1,l_{ij},\, \, 1\, \leq \, i\, <\,
j\, \leq \, n + 1} \nonumber \\ \left( \prod_{1\, \leq \, i\, \leq
\, n_{1} + 1,\,
 \, n_{1} + 1\, <\, j\, \leq \, n + 1} (- (a,a))^{- l_{\pi (i) \pi
(j)}} t_{m_{\pi (i) \pi (j)}\dot{m}_{\pi (i) \pi (j)}}^{l_{\pi (i)
\pi (j)}}(- \sqrt{- 1} \tilde{a}) \right) \times \nonumber \\ \left(
\prod_{1\, \leq \, i\, <\, j\, \leq \, n_{1} + 1,\, \, n_{1} + 1\, <
i\, <\, j\, \leq \, n + 1} \int_{0}^{\infty}  \mu_{\pi (i)\pi (j)}
d\mu_{\pi (i) \pi (j)} \right) \int d^{4(n + 1)}x\phi (x_{\pi
(1)},...,x_{\pi (n + 1)}) \times \nonumber \\
S_{m_{i},\dot{m}_{i},\, \, 1\, \leq \, i\, \leq \, n + 1 ;\, \,
m_{ij},\dot{m}_{ij},\, \, 1\, \leq \, i\, <\, j\, \leq \, n +
1}^{l_{i},\dot{l}_{i}, \, \, 1\, \leq \, i\, \leq \, n + 1;\, \,
l_{ij},\, \, 1\, \leq \, i\, <\, j\, \leq \, n + 1} \nonumber \\
(\pi ; A_{1},...,A_{N}; \mu_{\pi (i) \pi (j)}, 1 \leq i < j \leq
n_{1} + 1, n_{1} + 1 < i < j \leq n + 1)
 \times \nonumber \\ \prod_{1\, \leq
\, i\, <\, j\, \leq \, n_{1} + 1,\, \, n_{1} + 1\, <\, i\, <\, j\,
\leq \, n + 1} (2\pi ( - (x_{\pi (j)} - x_{\pi (i)},x_{\pi (j)} -
x_{\pi (i)}))^{q}) \times \nonumber \\ t_{m_{\pi (i) \pi
(j)}\dot{m}_{\pi (i) \pi (j)}}^{l_{\pi (i) \pi (j)}}(\sqrt{- 1}
 (\tilde{x}_{\pi (i)} - \tilde{x}_{\pi (j)})) ( - \mu_{\pi (i) \pi (j)}
(x_{\pi (j)} - x_{\pi (i)},x_{\pi (j)} - x_{\pi (i)}))^{- 1/2}
\times \nonumber \\ K_{1}(( - \mu_{\pi (i) \pi (j)} (x_{\pi (j)} -
x_{\pi (i)},x_{\pi (j)} - x_{\pi (i)}))^{1/2})
\end{eqnarray}
{\it where the function}
\begin{eqnarray}
\label{4.10} S_{m_{i},\dot{m}_{i},\, \, 1\, \leq \, i\, \leq \, n +
1 ;\, \, m_{ij},\dot{m}_{ij},\, \, 1\, \leq \, i\, <\, j\, \leq \, n
+ 1}^{l_{i},\dot{l}_{i}, \, \, 1\, \leq \, i\, \leq \, n + 1;\, \,
l_{ij},\, \, 1\, \leq \, i\, <\, j\, \leq \, n + 1} \nonumber \\
(\pi ; A_{1},...,A_{N}; \mu_{\pi (i) \pi (j)}, 1 \leq i < j \leq
n_{1} + 1, n_{1} + 1 < i < j \leq n + 1) = \nonumber \\ \left(
\prod_{1\, \leq \, i\, \leq \, n_{1} + 1,\, \, n_{1} + 1\, <\, j\,
\leq \, n + 1} \int_{0}^{\infty}  d\mu_{\pi (i) \pi (j)} \right)
\nonumber
\\ \hbox{as}(f)_{m_{i},\dot{m}_{i},\, \, 1\, \leq \, i\, \leq \, n +
1 ;\, \, m_{ij},\dot{m}_{ij},\, \, 1\, \leq \, i\, <\, j\, \leq \, n
+ 1}^{l_{i},\dot{l}_{i},\, \, 1\, \leq \, i\, \leq \, n + 1;\, \,
l_{ij},\, \, 1\, \leq \, i\, <\, j\, \leq \, n + 1} (\pi
;A_{1},...,A_{N};\mu_{ij}, 1 \leq i < j \leq n + 1)\times \nonumber
\\ \prod_{1\, \leq \, i\, \leq \, n_{1} + 1,\, \,
n_{1} + 1\, <\, j\, \leq \, n + 1} 2\pi (\mu_{\pi (i) \pi (j)})^{
1/2} K_{1}((\mu_{\pi (i) \pi (j)})^{1/2}).
\end{eqnarray}
{\it We use the same symbol} $\pi$ {\it for the number and a
permutation.}

\noindent {\it Proof.} Let an infinitely differentiable function
$\phi (x_{\pi (i)}, 1 \leq i \leq n + 1)$ have a compact support in
the domain $(x_{\pi (j)} - x_{\pi (i)},x_{\pi (j)} - x_{\pi (i)}) <
0$, $1 \leq i < j \leq n_{1} + 1$, $n_{1} + 1 < i < j \leq n + 1$.
Then in view of the relation (\ref{4.6}) for sufficiently large
number $t$ and for any vector $a \in {\bf R}^{4}$, $(a,a) < 0$, we
have
\begin{eqnarray}
\label{4.11} \int d^{4(n + 1)}x W_{m_{1},...,m_{n + 1};
\dot{m}_{1},..., \dot{m}_{n + 1}}^{l_{1} ,...,l_{n + 1};
\dot{l}_{1},..., \dot{l}_{n + 1}} (A_{1},...,A_{N};x_{2} -
x_{1},...,x_{n + 1} - x_{n}) \times \nonumber \\ \phi (x_{\pi (i)},
1 \leq i \leq n_{1} + 1; x_{\pi (i)} - ta, n_{1} + 1 < i \leq n + 1)
= \nonumber \\ \sum_{l_{ij} \in 1/2{\bf Z}_{+}, \, \, 1\,
 \leq \, i\, <\, j\, \leq \, n + 1} \sum_{m_{ij},\dot{m}_{ij} = -
l_{ij}, - l_{ij} + 1,...,l_{ij} - 1,l_{ij},\, \, 1\, \leq \, i\, <\,
j\, \leq \, n + 1} \nonumber \\ \left( \prod_{1\, \leq \, i\, <\,
j\, \leq \, n + 1}
\int_{0}^{\infty} \mu_{ij} d\mu_{ij} \right) \nonumber \\
f_{m_{i},\dot{m}_{i},\, \, 1\, \leq \, i\, \leq \, n + 1 ;\, \,
m_{ij},\dot{m}_{ij},\, \, 1\, \leq \, i\, <\, j\, \leq \, n +
1}^{l_{i},\dot{l}_{i},\, \, 1\, \leq \, i\, \leq \, n + 1;\, \,
l_{ij}\, \, 1\, \leq \, i\, <\, j\, \leq \, n + 1}
(A_{1},...,A_{N};\mu_{ij}, 1 \leq i < j \leq n + 1)\times \nonumber \\
\int d^{4(n + 1)}x\phi (x_{\pi (1)},...,x_{\pi (n + 1)}) G(a,\mu,
x)H(\mu, x), \nonumber \\ G(a,\mu, x) =  \prod_{1\, \leq \, i\, \leq
\, n_{1} + 1,\, \, n_{1} + 1\, <\, j\, \leq \, n + 1} (2\pi ( -
(x_{\pi (j)} - x_{\pi
(i)} + ta,x_{\pi (j)} - x_{\pi (i)} + ta))^{q}) \times \nonumber \\
t_{m_{\pi (i) \pi (j)}\dot{m}_{\pi (i) \pi (j)}}^{l_{\pi (i) \pi
(j)}}(\sqrt{- 1} (\tilde{x}_{\pi (i)} - \tilde{x}_{\pi (j)} -
t\tilde{a})) \times \nonumber \\ ( - \mu_{\pi (i) \pi (j)} (x_{\pi
(j)} - x_{\pi (i)} + ta,x_{\pi (j)} - x_{\pi (i)} + ta))^{-
1/2}\times \nonumber \\ K_{1}(( - \mu_{\pi (i) \pi (j)} (x_{\pi (j)}
- x_{\pi (i)} + ta,x_{\pi (j)} - x_{\pi (i)} + ta))^{1/2}),
\nonumber \\ H(\mu, x) = \prod_{1\, \leq \, i\, <\, j\, \leq \,
n_{1} + 1,\, \, n_{1} + 1\, <\, i\, <\, j\, \leq \, n + 1} (2\pi ( -
(x_{\pi (j)} - x_{\pi
(i)},x_{\pi (j)} - x_{\pi (i)}))^{q}) \times \nonumber \\
t_{m_{\pi (i) \pi (j)}\dot{m}_{\pi (i) \pi (j)}}^{l_{\pi (i) \pi
(j)}}(\sqrt{- 1} (\tilde{x}_{\pi (i)} - \tilde{x}_{\pi (j)})) ( -
\mu_{\pi (i) \pi (j)} (x_{\pi (j)} - x_{\pi
(i)},x_{\pi (j)} - x_{\pi (i)}))^{- 1/2}\times \nonumber \\
K_{1}(( - \mu_{\pi (i) \pi (j)} (x_{\pi (j)} - x_{\pi (i)},x_{\pi
(j)} - x_{\pi (i)}))^{1/2}).
\end{eqnarray}
The series in the right - hand side of the equality (\ref{4.11}) are
finite. If a vector $x \in {\bf R}^{4}$ lies in a compact set in the
domain $(x,x) < 0$, then in view of the relation (\ref{3.23}) the
function
$$
\mu (- \mu (x,x))^{- 1/2}K_{1}((- \mu (x,x))^{1/2})
$$
of the variable $\mu$ rapidly decreases on the positive semi - axis.
We change the number $t$ by the number $t(- (a,a))^{- 1/2}$. Now the
relations (\ref{4.7}), (\ref{4.11}) imply the relation (\ref{4.9}).
The lemma is proved.

Let the function (\ref{4.10}) be not zero only for the spins $l_{\pi
(i) \pi (j)} = 0$, $1 \leq i \leq n_{1} + 1$, $n_{1} + 1 < j \leq n
+ 1$ and for these spins
\begin{eqnarray}
\label{4.12} S_{m_{i},\dot{m}_{i},\, \, 1\, \leq \, i\, \leq \, n +
1 ;\, \, m_{ij},\dot{m}_{ij},\, \, 1\, \leq \, i\, <\, j\, \leq \, n
+ 1}^{l_{i},\dot{l}_{i}, \, \, 1\, \leq \, i\, \leq \, n + 1;\, \,
l_{ij}\, \, 1\, \leq \, i\, <\, j\, \leq \, n + 1} \nonumber \\ (\pi
; A_{1},...,A_{N}; \mu_{\pi (i) \pi (j)}, 1 \leq i < j \leq n_{1} +
1, n_{1} + 1 < i < j \leq n + 1) = \nonumber \\
f_{m_{\pi (i)},\dot{m}_{\pi (i)},\, \, 1\, \leq \, i\, \leq \, n_{1}
+ 1 ;\, \, m_{\pi (i) \pi (j)},\dot{m}_{\pi (i) \pi (j)},\, \, 1\,
\leq \, i\, <\, j\, \leq \, n_{1} + 1}^{l_{\pi (i)},\dot{l}_{\pi
(i)},\, \, 1\, \leq \, i\, \leq \, n_{1} + 1;\, \,
l_{\pi (i) \pi (j)}\, \, 1\, \leq \, i\, <\, j\, \leq \, n_{1} + 1} \nonumber \\
(\pi (A_{1}^{(1)}),...,\pi (A_{N_{1}}^{(1)});\mu_{\pi (i) \pi (j)}, 1 \leq i < j \leq n_{1} + 1) \times \nonumber \\
f_{m_{\pi (i)},\dot{m}_{\pi (i)},\, \, n_{1} + 1\, \leq \, i\, \leq
\, n + 1 ;\, \, m_{\pi (i) \pi (j)},\dot{m}_{\pi (i) \pi (j)}, \, \,
n_{1} + 1\, \leq \, i\, <\, j\, \leq \, n + 1}^{l_{\pi (i)},
\dot{l}_{\pi (i)},\, \, n_{1} + 1\, \leq \, i\, \leq \, n + 1;\, \,
l_{\pi (i) \pi (j)}\, \, n_{1} + 1\, \leq \, i\, <\, j\, \leq \, n + 1} \nonumber \\
(\pi (A_{1}^{(2)}),...,\pi (A_{N_{2}}^{(2)});\mu_{\pi (i) \pi (j)},
n_{1} + 1 < i < j \leq n + 1)
\end{eqnarray}
The substitution of the function (\ref{4.12}) into the relation
(\ref{4.9}) yields the relation of the type (\ref{4.1}).

\end{document}